\documentclass[journal]{IEEEtran}
\usepackage{amsmath}
\usepackage{amssymb}
\usepackage{graphicx}
\usepackage{algpseudocode}
\usepackage[margin=0.75in]{geometry}
\usepackage{cite}
\usepackage{inputenc}
\usepackage{amsthm}
\theoremstyle{definition}

\newtheorem{theorem}{Theorem}
\newtheorem{proposition}{Proposition}
\newtheorem{lemma}{Lemma}

\begin{document}

\title{Optimization of Heterogeneous Coded Caching} 

\author{Alexander Michael Daniel and Wei Yu
\thanks{Manuscript prepared %submitted to IEEE Transactions on Information Theory 
on \today. This work is supported by Natural Science
and Engineering Research Council (NSERC).
The authors are with
the Edward S. Rogers Sr.  Department of Electrical and Computer
Engineering, University of Toronto, Toronto, ON M5S 3G4, Canada
(e-mails: alex.daniel@mail.utoronto.ca, weiyu@ece.utoronto.ca).}%
}

% The paper headers
%\markboth{}
\markboth{Daniel, Yu: Optimization of Heterogeneous Coded Caching}{}

% use for special paper notices
%\IEEEspecialpapernotice{(Invited Paper)}

% make the title area
\maketitle

%Abstract
\begin{abstract}
This paper aims to provide an optimization framework for coded caching that
accounts for various heterogeneous aspects of practical systems. %This framework is first used to develop 
An optimization theoretic perspective on the seminal work on the fundamental
limits of caching by Maddah Ali and Niesen is first developed, whereas it is
proved that the coded caching scheme presented in that work is the optimal
scheme among a large, non-trivial family of possible caching schemes. The
optimization framework is then used to develop a coded caching scheme capable of
handling simultaneous non-uniform file length, non-uniform file popularity, and
non-uniform user cache size. Although the resulting full optimization problem
scales exponentially with the problem size, this paper shows that tractable
simplifications of the problem that scale as a polynomial function of the
problem size can still perform well compared to the original problem. By
considering these heterogeneities both individually and in conjunction with 
one another, insights into their interactions and influence on optimal cache
content are obtained.  
\end{abstract}


\section{Introduction}

\subsection{Background}
%\cite{6736746} 5 Disruptive technologies for 5G
Caching technologies stand poised to make an important contribution to future
5G cellular networks \cite{6736746}.  One such technique is \emph{coded
caching}, which, roughly speaking, is the idea of using carefully designed user
cache content to enable content delivery via coded multicast transmissions.
(The cached contents themselves are uncoded.)
First introduced by Maddah-Ali and Niesen in \cite{6620392, 6763007}, coded
caching has since been the subject of a great number of studies seeking to
extend the original scheme into more practical scenarios. The coded caching scheme of
\cite{6620392, 6763007} is developed for a system in which a central server has
complete knowledge of user numbers and identities, users have identical cache
sizes and make a single download request, transmission occurs over an error
free link, and files are of equal length and popularity. Subsequent work has
since extended the coded caching idea to the decentralized system 
\cite{6807823}, and to systems with non-uniform file popularity \cite{6849235,7782760, 7904696,
6933485, 2016arXiv160505026C, 2016arXiv160905836H, 2016arXiv160905831H,
7218445,7865913, 6874794, 7282746,7308972, 2017arXiv170707146J}, non-uniform
file length \cite{7282743, 7523930}, multiple user requests
\cite{7282744,7843674,2015arXiv151107542J,2014arXiv1402.4572J}, non-uniform
cache size \cite{2015arXiv150401123W,2016arXiv161101579A,7869142,7925535}, and
non-uniform channel quality \cite{2016arXiv160502317S, 2015arXiv150501016T,
7541613,7593135, 2016arXiv160608253Z, 2017arXiv170208044S, 7558129,
2017arXiv170205454A, 2017arXiv170107730D, 2017arXiv170202179G,
2016arXiv161104853Z, 2015arXiv150401452H, 7485891}. 

The aforementioned works typically either discuss how the scheme of
\cite{6620392, 6763007} should be modified to accommodate the considered
heterogeneity, or develop an entirely new scheme that enables coded multicast
transmissions while accommodating the aforementioned heterogeneity. 
Most of these papers, however, consider only one type of non-uniformity.
While this is sensible from the viewpoint of understanding how each
heterogeneity affects coded caching systems by itself, practical systems would
have to account multiple types of non-uniform parameters. Moreover, we cannot
in general expect the effects of these non-uniformities to be additive. It is
thus important to consider combinations of heterogeneities, which, to the best
of our knowledge, only a few recent works have started to explore: the recent
work \cite{7523930} examines the achievable rate region for a system serving
two users with two files, where the user cache sizes and file lengths are not
necessarily uniform, while some of the aforementioned work on caching with
non-uniform channel quality exploits heterogeneous cache size to rectify
disparities in channel quality (see e.g. \cite{2017arXiv170208044S} and
references therein). 

\subsection{Main Contributions}

This paper proposes an optimization theoretic framework to design 
caching schemes capable of accommodating non-uniform file length,
non-uniform file popularity, and non-uniform user cache size at the
same time. More specifically, we design a caching scheme that uses a
generalized version of the transmission scheme of \cite{6807823},
paired with an optimization problem designed to yield the optimally
coded content. This optimization problem, although convex, has the
number of variables, constraints, and objective function terms that
scale exponentially with the problem size, so subsequently this paper
develops high-quality simplifications that scale polynomially with
the problem size, yet perform well compared to the original problem. 
%, but have a number of variables, constraints, and objective function terms that scales as a polynomial function of the system parameters. These optimization problems not
The proposed optimization approach only yields numerical answers
corresponding to the optimized caching schemes, but also generate
practical insight into the problems considered. 
%The primary contribution of this paper is twofold. First, this paper contains the study of, and the development of caching schemes for, certain practical systems that, to the best of our knowledge, have yet to be considered in the literature. The aforementioned exponentially scaled optimization problem appears to be the first caching scheme to accommodate simultaneous non-uniformity in cache size, file size, and/or popularity, but also the first caching scheme to accommodate any two of those heterogeneities for a general number of users and files.
%
%The second significant contribution is the further development of an optimization approach to coded caching problems that is likely to be useful beyond the cases considered in this work. 

Optimization approaches have been used in the past for content placement for 
%While it is relatively common to optimize cache content in studies of general caching for wireless networks (e.g. in the well-known work on ``
femtocaching systems \cite{6195469, 6600983}, but its use in the coded
caching context has only appeared recently: %whether to design cache content or otherwise. 
for instance, \cite{2016arXiv161100024T} uses an information-theory
based optimization problem to help characterize the achievable rate
region for certain numbers of users and files in the case where all
other systems parameters are uniform. More related is the recent work
\cite{7925535} in which an optimization framework similar to the one
used here is employed to develop a caching scheme in the case of
non-uniform cache size.  Crucially, both approaches design cache
content in terms of the subsets of users who have cached the content
(see Section \ref{AOTPOFLC}). While we directly express our
transmission scheme in terms of that cache content, the approach in
\cite{7925535} is to further design two sets of variables through
which the transmission scheme is expressed. Moreover, the framework
used in \cite{7925535} does not distinguish between different files
beyond designing a scheme for the worst-case scenario where users
request different files. This prevents their framework from addressing
file heterogeneity, whereas ours allows for it. Finally, while the
optimization problem in \cite{7925535} suffers from the same
exponential scaling problem that the general problem here has, 
they do not present any tractable methods of solution like we do 
here.  %Nevertheless, we take \cite{7925535} as further evidence that optimization frameworks like the ones presented there and here are of practical utility.

During the preparation of this paper, we became aware of independent
work \cite{2017arXiv170707146J}, which uses an optimization framework
essentially the same as the one proposed here to study the case of
non-uniform file popularity. While independently and simultaneously
developed, a number of results in \cite{2017arXiv170707146J} are
echoed in this paper. Specifically, both works develop the same
exponential-order general optimization problem; then a simplified
polynomial-order optimization problem is developed for the
non-uniform popularity case (Section \ref{PF1} of this paper), 
although the exact formulations are different.
Moreover, both works show that in the special case of uniform
popularity, the caching and delivery scheme of \cite{6620392,6763007}
is the optimal solution (Section \ref{AOTPOFLC} of this paper);
however, the proof in \cite{2017arXiv170707146J} is quite involved,
while a much simpler proof is presented here. 
%they differ in
%the particular details, and so we include our proof in this paper to
%provide an alternate perspective. 
Ultimately, the focus of
\cite{2017arXiv170707146J} is to study the case of non-uniform
popularity in great depth, while here, it is only considered as an
intermediate step towards the study of the interactions of several
heterogeneities at the same time. Thus, despite the similarities, both
papers develop many unique insights of practical significance. Indeed,
the results of \cite{7925535} and \cite{2017arXiv170707146J} taken
jointly with the results of this paper suggest that the optimization
framework common to all three works is likely to be a useful one for
coded caching.

\subsection{Notation}
The notation $[a:b]$ is used as
shorthand for the set of consecutive integers $\{a, a+1, \dots, b-1, b\}$, and $[b]$ is used as an
abbreviation of $[1:b]$. The symbol $\oplus$ is used to denote the
bitwise ``XOR'' operation between two or more files (i.e. strings of
bits). Both $l$ and $W^{(l)}$ are used to refer to the $l$-th file
under consideration. For an arbitrary file $W^{(n)}$, $|W^{(n)}|$
refers to the length of the file, and $W^{(l)}_{\mathcal{S}}$, called a ``subfile'' of file $W^{(l)}$, refers to the portion of file $l$ stored exclusively on the caches of the users in the set $\mathcal{S}$. For notational convenience, notation of the form $W^{(l)}_{123}$ is used instead of $W^{(i)}_{\{1,2,3\}}$ (for the $\mathcal{S} = \{1,2,3\}$ case in this example), returning only to the latter notation if necessary to resolve ambiguity. For a set $\mathcal{S}$, $\mathcal{P}(\mathcal{S})$ refers to the power set of $\mathcal{S}$.
For a real number $t$, $\lfloor \cdot \rfloor$ and $\lceil \cdot
\rceil$ denote the floor and ceiling functions, respectively.

We define the binomial coefficient $\binom{n}{k}$ in the usual way for $0 \leq k \leq n$, i.e. $\binom{n}{k} = n!/(k!(n-k!))$, but for $n < 0$ or $k > n$, we take $\binom{n}{k}=0$. Moreover, the notation of the so-called ``multinomial coefficient" is used, defined as:
\begin{equation}
\binom{n}{k_1,k_2, \dots, k_m} =  \frac{n!}{k_1!k_2!\dots k_m!},\nonumber
\end{equation}
where $k_1 + k_2 + \dots +k_m = n$. We use the notation $\sum_{i = a}^{b}n_i$ in the usual way when $a \leq b$, and take $\sum_{i = a}^{b}n_i = 0$ identically if $a > b$. More generally, a sum over an empty set of indices is taken to equal zero. Finally, if a sum over an empty set is raised to the power of zero, we take the resulting value to be 1; this is simply used for notational convenience.

\subsection{Organization}

The organization of the rest of the paper is as follows. Section
\ref{Review} introduces the optimization framework used in this paper
and Section \ref{AOTPOFLC} uses this framework to provide a new
interpretation and understanding of \cite{6620392, 6763007,6807823} by
showing how they represent feasible points in the optimization
problem. Sections \ref{NU1}-\ref{NU3} use the framework to develop
tractable simplifications of the original problem for different types
and combinations of non-uniformities. Section \ref{Conclusion}
concludes the paper with a summary and discussion of main results.

\section{Optimization Perspective on Coded Caching}

\label{Review}

We begin by providing an optimization perspective on the 
%For the sake of brevity, the reader is assumed to be familiar with the
coded caching schemes of Maddah-Ali and Niesen for both the
centralized \cite{6620392, 6763007} and the decentralized \cite{6807823}
cases. The transmission scenario consists of a server with a
set of $N$ files, $\mathcal{F} = \{1,2,\dots, N\} = [N]$, serving a
set of $K$ users, $\mathcal{U} = \{1, 2, \dots, K\} = [K]$, over a
shared, error-free link. %Instead of the system with uniform parameter as considered in \cite{6620392, 6763007, 6807823}, 
For full generality, we allow each file $l$ to have a distinct length
of $F_l$ bits and a distinct probability $p_l$ of being requested, and
allow each user $k$ to have arbitrary cache size of $M_k$ bits. 

Central to the coded caching scheme of \cite{6620392, 6763007,
6807823} is the partitioning of each file $l \in \mathcal{F}$ into
subfiles $W^{(l)}_{\mathcal{S}}$, indexed by all subsets of
$\mathcal{S} \subseteq \mathcal{U}$. (No two subfiles have a non-empty
intersection; together the subfiles jointly reconstruct the original
file.) In the caching phase, each user $k$ caches 
%all the subfiles whose indices contain the user index. 
\begin{equation}
\bigcup_{l, \mathcal{S}\in \mathcal{P}(\mathcal{U}\setminus\{ k\})}
W^{(l)}_{\mathcal{S}\cup k}
\end{equation}
without coding.  In the content delivery phase, given the set of user
requests $\mathbf{d} = [d_1, \dots, d_K]$, where $d_k$ denotes the
index of the file requested by user $k$, the server transmits
\begin{equation}
\bigoplus_{k \in \mathcal{S}} W^{(d_k)}_{\mathcal{S}\setminus \{k\} }  \label{coded}
\end{equation}
over the shared link with zero-padding if necessary, for each $\mathcal{S} \in
\mathcal{P}(\mathcal{U})\setminus \emptyset$, %where $W^{(d_k)}_{\mathcal{S}\setminus \{k\}}$ denotes subfiles that partition file $d_k$ and indexed by the set $\mathcal{S} \setminus \{k\}$, 
so that together with uncoded content stored a priori in each user's 
local cache, all users are guaranteed to be able to reconstruct the
entirety of their requested files. 
%Both the centralized scheme looks like, in a sense, a ``special case'' of the decentralized scheme: the same transmission scheme can be used for both.  

Any coded caching scheme (with coded transmission and uncoded cache) 
can be described using this ``subset partitioning'' representation
\cite{6807823}.  Different caching strategies differ in their
partitioning of the subfiles. But instead of thinking of the above as
a way of \emph{labelling} the cache contents, we can also use this to
\emph{design} the cache content, and regard the size of each subfile
$W^{(l)}_{\mathcal{S}}$ as design variables. 
%\footnote{We acknowledge that the centralized and decentralized schemes are in fact meant to address different problems.  Nevertheless, it is possible to think of the decentralized scheme as representing one possible solution to the caching problem in the centralized case, where the server has delivered the cache content randomly rather than through careful design} 
%In fact, both of these schemes can be taken to represent to special cases of a broad class of caching schemes defined above. 
%\footnote{The typical use of the word ``partition'' in set theory requires each subset in the partition to be non-empty, but we use the word here without that condition.}  
Consequently, instead of designing cache contents by
assigning discrete bits to sets,  the problem can be simplified by
designing only the sizes of the subfiles: if we have
$|W^{(l)}_{\emptyset}| = b^{(l)}_{\emptyset}, |W^{(l)}_{1}| =
b^{(l)}_1, \dots ,|W^{(l)}_{\mathcal{U}}| = b^{(l)}_{\mathcal{U}}$,
then the first $b^{(l)}_{\emptyset}F_l$ bits of $W^{(l)}$ are assigned
to $W^{(l)}_{\emptyset}$, the next $b^{(l)}_1F_l$ bits to
$W^{(l)}_{1}$, and so on, assigning the final
$b^{(l)}_{\mathcal{U}}F_l$ bits to $W^{(l)}_{\mathcal{U}}$. (Formally,
this requires that $b^{(l)}_{\mathcal{S}} \in \{0, 1/F_l, \dots
(F_l-1)/F_l, 1\}$, but for large $F_l$, this can be relaxed to
$b^{(l)}_{\mathcal{S}} \in [0,1]$ without any significant loss.)

For example, consider the centralized scenario wherein the numbers 
and identities of users are known in advance, so the server has the
ability to design the cache content of each user.  The coded caching
scheme of \cite{6620392, 6763007} (assuming uniform file length, cache
size and uniform popularity) sets $|W_{\mathcal{S}}|$ to be non-zero 
for only the $\mathcal{S}$'s with $|\mathcal{S}|=t=KM/N$.
In the decentralized setting of \cite{6807823}, all subfiles have
non-zero sizes due to the use of random cache content. More
generally, $|W^{(l)}_{\mathcal{S}}|$ can be explicitly designed.

The design of coded caching in this way imposes some natural
constraints on the subfile sizes. First, the subfiles together must 
contain the entire file, i.e.  
\begin{equation}
\sum_{\mathcal{S}\in \mathcal{P}(\mathcal{U})} |W^{(l)}_{\mathcal{S}}| = F_l, \;\; \forall l \in \mathcal{F}. \label{firstconstraint}
\end{equation}
Second, denoting the amount of cache dedicated to file $l$ by user $k$ as $\mu_{k,l}$, the amount of a file cached by a user is expressed as
\begin{equation}
\sum_{\mathcal{S}\in \mathcal{P}(\mathcal{U}\setminus\{ k\})} |W^{(l)}_{\mathcal{S}\cup k}| \leq \mu_{k,l}, \;\; \forall k \in \mathcal{U}, \forall l \in \mathcal{F} \label{secondconstraint},
\end{equation}
where
\begin{equation}
\sum_{l = 1}^N \mu_{k,l} = M_k, \forall k \in [K].
\end{equation}
Finally, the subfiles cannot have a negative size:
\begin{equation}
|W^{(l)}_{\mathcal{S}}| \geq 0 , \quad \forall \mathcal{S}\in
\mathcal{P}(\mathcal{U}), \quad \forall l \in \mathcal{F}. \label{thirdconstraint} 
\end{equation}

In the transmission defined in \eqref{coded},
zero-padding is neeeded whenever the subfiles do not have the same length, 
% for all $\mathcal{S} \in \mathcal{P}(\mathcal{U})$, with Specifically, to send, say, $W^{(1)}$ and $W^{(2)}$, with $|W^{(1)}| > |W^{(2)}|$, form $\hat{W}^{(2)} = W^{(2)} \cup \{ 0, 0, \dots, 0\}$ such that $|\hat{W}^{(2)}| = |W^{(1)}|$. A user who has cached $W^{(2)}$ and wishes to decode $W^{(1)}$ can still do so in the usual way, but 
so the length of a single transmission is determined
by the largest subfile in the transmission. For a vector of user
requests $\mathbf{d}$, the number of bits sent to satisfy user
requests given in $\mathbf{d}$ is thus 
\begin{equation}
R_\mathbf{d} = \sum_{\mathcal{S}\in \mathcal{P}(\mathcal{U})\setminus
\emptyset} \max_{k \in \mathcal{S}} \left\{|W^{(d_k)}_{\mathcal{S}\setminus \{k\} }| 
\right\} \label{objfunc}
\end{equation}

The set of choices of $|W^{(l)}_{\mathcal{S}}|$
define a broad family of caching schemes. %each one obtained by choosing a different set of feasible $|W^{(l)}_{\mathcal{S}}|$ values. 
To find the most efficient caching strategy among this family of
schemes that minimize the expected delivery rate over all demand
requests, we can formulate the following optimization problem:
\begin{align}
\mbox{minimize} \;\;\; &\mathbb{E}[R_{\mathbf{d}}] = \sum_{\mathbf{d}
\in \mathcal{F}^K}p(\mathbf{d})\sum_{\mathcal{S}\in
\mathcal{P}(\mathcal{U})\setminus \emptyset} \max_{k \in \mathcal{S}}
\left\{|W^{(d_k)}_{\mathcal{S}\setminus \{k\} }| \right\}  \label{generalorig}\\
\mbox{subject to} \;\;\; & \text{ \eqref{firstconstraint}-\eqref{thirdconstraint}.} \label{generalconstraint}
\end{align}

We note before continuing that the transmission scheme described above
is the same for all possible user requests, which may be suboptimal if
there is repetition in the users' requests, i.e.  the same file is
requested by more than one user. However, it can readily be shown that
the probability of each user requesting a distinct file goes to one as
$N \rightarrow \infty$ of a fixed $K$. Since it is likely that $N \gg
K$ in practice, the optimization formulation \eqref{generalorig} can
be therefore be used without fear of significant loss. 

The optimization problem \eqref{generalorig}-\eqref{generalconstraint}
is convex in the $|W_{\mathcal{S}}|$ variables. % since the constraints are linear and the maximum function is a convex function of its argument. 
However,
there are $N2^K$ variables, $N^K(2^K-1)$ summands in the objective
function, and $N2^K+KN+N+K$ constraints, making it impractical
to solve for large-scale problems. The rest of this paper is dedicated to
developing simplifications of
\eqref{generalorig}-\eqref{generalconstraint} that allow for high
quality (and even optimal) solution while maintaining a tractable
problem size.

\section{Homogeneous Coded Caching}
\label{AOTPOFLC}

Consider the special case of problem
\eqref{generalorig}-\eqref{generalconstraint} with uniform
file lengths, $F_l = F, \forall l \in [N]$, uniform file popularities,
$p_l = 1/N, \forall l \in [N]$, and uniform cache sizes $M_k = M,
\forall k \in [K]$; this is the same system originally considered in
\cite{6620392, 6763007}. The symmetry of the resulting problem can be
exploited to reduce the computational complexity of optimization.
Specifically, define $v_j$ such that $v_j = |W_{\mathcal{S}}|$ for all
$\mathcal{S}$ such that $|\mathcal{S}| = j$ and for all files $W$; this reduces the number of optimization variables from an exponential number in $K$ to a linear number in $K$.
Since the file length, file popularity and cache size are all homogeneous,
there is symmetry across both the users and the files, and so we would
expect the solution to \eqref{generalorig}-\eqref{generalconstraint}
for this uniform case to have this form, i.e., any two subfiles have
the same size if their respective user sets are the same size. % regardless of which file the subfiles are taken from. By the symmetry across both users and files, we would expect this to be the case for the 
%with uniform file lengths, file popularities, user cache sizes. Additionally, the problem is simplified drastically by rewriting in terms of the $v_j$ variables.
%This class of caching schemes contains both the schemes of \cite{6620392, 6763007} and \cite{6807823}; the former scheme is obtained by setting, for all $W$, $|W_{S}| = 1/\binom{K}{t}$ if $|\mathcal{S}| = t$, and $|W_{\mathcal{S}}| = 0$ otherwise, whereas the latter scheme is obtained with $|W_{\mathcal{S}} | = (M/N)^{|\mathcal{S}|}(1-M/N)^{K - |\mathcal{S}|}F$. 
%Since the subfiles indexed by equal-sized user subsets are the same size, 
The summands of the objective function now simplify as 
\begin{equation}
\max_{k \in \mathcal{S}} \left\{|W^{(k)}_{\mathcal{S}\setminus \{k\}
}| \right\} = |W^{(k')}_{\mathcal{S}\setminus \{k'\} }| = v_{|\mathcal{S}|-1} \nonumber
\end{equation}
where $k'$ is any user in the set $\mathcal{S}\setminus \{k\}$, because each subfile in a given transmission is the same size.
Since the files are equally popular, every request vector is equally likely, and since every
transmission scheme requires the same number of bits to satisfy the
requests, the average does not need to be taken across all the $N^K$ possible demands as in
\eqref{generalorig}. Finally, since $v_jF$ bits are sent to any subset
of $j+1$ users, and there are $\binom{K}{j+1}$ subsets of $j+1$ users,
the objective function becomes
\begin{equation}
\mathbb{E}[R_{\mathbf{d}}] = \sum^{K-1}_{j=0} \binom{K}{j+1} v_{j}F, \label{2ndobjfunc}
\end{equation}
Note that the number of terms in the objective function now scales linearly in $K$ instead of exponentially in $K$, because each term in \eqref{2ndobjfunc} accounts for a combinatorial number of transmissions.

The constraints simplify as well. Since all files are of equal length, only one file reconstruction constraint is required. Then, because there are $\binom{K}{j}$ subsets of size $j$, the file reconstruction constraint becomes
\begin{equation}
\sum_{j=0}^{K}\binom{K}{j}v_j = F \label{2ndfirstconstraint}.
\end{equation} 
Moreover, since the cache sizes are uniform, only one cache constraint is needed, and since all file are homogeneous, an equal amount of memory is allocated to each. The cache constraint thus simplifies to 
\begin{equation}
\sum_{j=1}^{K} \binom{K-1}{j-1}v_j \leq MF/N, \label{2ndsecondconstraint}
\end{equation}
because there are $\binom{K-1}{j-1}$ subsets of size $j$ that contain
the index $k$. %Finally, it is straightforward to see that the final constraint simplifies to \begin{equation} v_j \geq 0 , \forall j \in \mathcal{U} \label{2ndthirdconstraint}.  \end{equation}
As a final step, we follow \cite{6620392, 6763007} and normalize the
file length 
$F = 1$ in \eqref{2ndobjfunc}-\eqref{2ndsecondconstraint}. This yields
the following linear programming problem with $K+1$ variables, $K+3$
constraints, and $K$ terms in the objective function:
%which obtains the optimal cache content among the broad family of caching schemes discussed earlier: 
\begin{align}
\mbox{minimize} \;\;\; & \sum^{K-1}_{j=0} \binom{K}{j+1} v_{j} \label{orig2}\\
\mbox{subject to} \;\;\; & \sum_{j=0}^{K}\binom{K}{j}v_j = 1, \label{12} \\
	       & \sum_{j=1}^{K} \binom{K-1}{j-1}v_j \leq M/N,  \label{22} \\
	       & v_j \geq 0 , \;\; \forall j \in \mathcal{U}. \label{32} 
\end{align}
Note that the reduction to an optimization problem that scales as a
linear function of $K$ and as a constant function of $N$ is possible
because of the symmetry created by the uniform file length, file
popularity, and cache size. Later in this paper, cases when one or
more of these parameters are non-uniform are treated; but the reduction
from the exponential order will no longer be without loss of generality.

%of \eqref{generalorig}-\eqref{generalconstraint} can still be reduced in these cases, it will not be to the degree seen here.

Note also that the schemes of \cite{6620392, 6763007} and \cite{6807823}
are all feasible points of this problem. In particular, assuming that 
$t = KM/N$ is an integer, the caching scheme of \cite{6620392, 6763007} 
sets $v_t = 1/\binom{K}{t}$ and $v_j = 0$ if $j \neq t$, while the
decentralized scheme of \cite{6807823} sets the variables to be of the 
form $v_j = (M/N)^j(1-M/N)^{K-j}$ for all $j$. For the non-integer $t$
case, a similar scheme is also stated in \cite{6620392, 6763007}.
The following theorem
shows that the caching scheme of \cite{6620392, 6763007} is, in fact,
the optimal scheme among the broad family of schemes discussed above.
The theorem works for the cases of integer or non-integer $t$. 
We note that while a similar result for the integer $t$ case
exists in \cite{2017arXiv170707146J}, the proof used here uses a novel
reformulation over the probability simplex. It is considerably simpler
and lends additional insight into the problem.

\begin{theorem}
\label{biggie}
Assume $M \le N$. The unique, optimal solution to \eqref{orig2}-\eqref{32} is:
\begin{itemize}
\item For $t = KM/N \in \mathbb{Z}$:
\begin{align}
 v_j^* = & \label{firstpart}
 \begin{cases}
1/\binom{K}{t} & \text{if } j = t, \\
0 & \text{if } j \neq t
\end{cases}
\end{align}
\item For $t = KM/N \notin \mathbb{Z}$: 
\begin{align}
v_j^* = & \label{secondpart}
\begin{cases}
s/\binom{K}{\lfloor t \rfloor} &\text{if } j = \lfloor t \rfloor \\
(1-s)/\binom{K}{\lceil t \rceil} &\text{if } j = \lceil t \rceil \\
0 & \text{else}
\end{cases}
\end{align}
where $s = \lceil t \rceil - t$. 
\end{itemize}
\end{theorem}

\begin{IEEEproof}
%\begin{theorem}
%\label{biggie}
%Assuming that $t = KM/N$ is an integer, % \in \{ 1, 2, \dots, K-1\}$, 
%a primal/dual optimal point for the problem \eqref{orig2}-\eqref{32} is $(\mathbf{v}^*,\mathbf{\mu}^*,\nu^*,\lambda^*)$, where 
%\begin{align}
%v_j^* = &
%\begin{cases}
%0 & \text{if } j \neq t, \\
%\frac{1}{\binom{K}{t}} & \text{if } j = t
%\end{cases}, \label{optimal} \\
%\mu_j^* = &
%\begin{cases}
%0 & \text{if } j = t, \\
%\binom{K}{j+1} + \lambda^*\binom{K-1}{j-1} + \nu^*\binom{K}{j} & \text{if } j \neq t
%\end{cases}, \label{optimal2} \\
%\nu^* = &-\frac{M}{N}\lambda^* - \frac{K(1-M/N)}{KM/N+1},
%\end{align}
%and $\lambda^*$ is any point satisfying the following two inequalities: 
%\begin{align}
 %\lambda^* \geq &\,\,
%\max\left\{\frac{K(1+K)}{\left(\frac{KM}{N}+1\right)\left(\frac{KM}{N}+2\right)},\frac{K}{\frac{KM}{N}+1}\right\} \\
 %\lambda^* \leq &\,\, \frac{K(1+K)}{\left(\frac{KM}{N}+1\right)\left(\frac{KM}{N}\right)}, \nonumber
%\end{align}
%where the first term in the maximum is selected if and only if $K \geq 1/(1-M/N)$. Such a point $\lambda^*$ always exists. 
%\end{theorem}
%The proof is provided in Appendix \ref{A}. 
%As mentioned earlier, the same result has also been obtained in the
%independent work \cite{2017arXiv170707146J}.
%While rigorous, it does not provide intuition as to why \eqref{optimal}-\eqref{optimal2} is indeed the optimal solution to the primal problem. A less rigorous but 
%An intuitive way to see this result is to 
%Before the proof of Theorem \ref{biggie} is stated, consider the 
First, we make a change of variables: $a_j = v_j/\binom{K}{j}$; after some mild algebra, the original
problem \eqref{orig2}-\eqref{32} can be reformulated as:
\begin{align}
\mbox{minimize} \;\;\; & \sum^{K}_{j=0} \frac{K-j}{j+1} a_{j} \label{orig3}\\
\mbox{subject to} \;\;\; & \sum_{j=0}^{K} ja_j \leq t  \label{23} \\
	       & \sum_{j=0}^{K} a_j = 1 \label{13} \\
	       & a_j \geq 0 , \quad \forall j \in [0:K]. \label{33} 
\end{align}
In this form, the optimization problem is more easily understood.
Constraints \eqref{13} and \eqref{33} restrict the feasible space to
the probability simplex. The key features of the formulation is that 
the coefficients of the objective function $\frac{K-j}{j+1}$ are
decreasing in $j$ with decreasing second differences, while 
the cache constraint \eqref{23} now has coefficients that increase 
linearly in $j$. Intuitively, the optimal solution would
involve placing as much ``probability mass'' in high-$j$ $a_j$
variables in order to lower the objective function, but not so high as to
violate the cache constraint. This results in the optimal $a_j$ to
concentrate around at most two consecutive $j$'s close to $t$.

To complete the argument, we first observe that at the
optimal solution $\mathbf{a}^*$ of \eqref{orig3}-\eqref{33}, the cache
constraint \eqref{23} must be tight, when $M\le N$. This is because
if it were not, one can always shift some of the weight of $a_j$ to a
higher indexed $a_{j+1}$ without violating the constraints while
lowering the objective function.  In the practical context, this means
that the optimal caching strategy does not waste any cache space. 

Second, for a similar reason, we can show that 
%\begin{lemma}
%\label{prooflemma2}
if a feasible solution $\mathbf{a} = [a_0, \dots , a_K]^T$ to
\eqref{orig3}-\eqref{33} has two non-zero variables $a_{i_1} \neq 0$
and $a_{i_2} \neq 0 $ such that $i_2 - i_1 \geq 2$, then
$\mathbf{a}$ cannot be an optimal solution of \eqref{orig3}-\eqref{33}.
%\end{lemma}
%\emph{Proof of Lemma \ref{prooflemma2}:} Like Lemma \ref{prooflemma2}, we will show that there always exists a better feasible solution than $\mathbf{a}$, which we denote by $\bar{\mathbf{a}} = [\bar{a}_0, \dots, \bar{a}_K]^T$. Without loss of generality, assume $|i_2 - i_1| > 2$ for convenience, although the proof technique is the same if $|i_2 - i_1| = 2$, and assume that $i_2 > i_1$. For some small $\Delta > 0$, set $\bar{a}_{i_1} = a_{i_1}-\Delta$, $\bar{a}_{i_1+1} = a_{i_1+1}+ \Delta$, $\bar{a}_{i_2} = a_{i_2}-\Delta$, $\bar{a}_{i_2-1} = a_{i_2-1}+\Delta$, and $\bar{a}_{j} = a_{j}$ for all other $j$ values. The only restriction on the size of $\Delta$ is that it be small enough such that $\bar{a}_{i_1}$ and $\bar{a}_{i_2}$ are still positive. And since $\Delta$ is being subtracted and added in equal measure, the point $\bar{\mathbf{a}}$ remains on the probability simplex.
This is because for any such $\bf{a}$, we can always construct a
better feasible solution $\bar{\mathbf{a}} =
[\bar{a}_0, \dots, \bar{a}_K]^T$ in the following way. 
%Without loss of generality, assume for convenience, although the proof technique is the same if $|i_2 - i_1| = 2$, and assume that $i_2 > i_1$. For some small $\Delta > 0$, set 
For some small $\Delta$, set
$\bar{a}_{i_1} = a_{i_1}-\Delta$, $\bar{a}_{i_1+1} = a_{i_1+1}+ \Delta$, $\bar{a}_{i_2}
= a_{i_2}-\Delta$, $\bar{a}_{i_2-1} = a_{i_2-1}+\Delta$, and
$\bar{a}_{j} = a_{j}$ for all other $j$ values. 
(If $i_1+1=i_2-1$, then set $\bar{a}_{i_2-1} = a_{i_2-1}+2\Delta$.) 
%The only restriction on the size of $\Delta$ is that it be small enough such that $\bar{a}_{i_1}$ and $\bar{a}_{i_2}$ are still positive. And since $\Delta$ is being subtracted and added in equal measure, the point 
Note that such $\bar{\mathbf{a}}$ remains on the probability simplex;
the cache constraint remains satisfied as 
$\sum_{j=0}^{K} j\bar{a}_j = \sum_{j=0}^{K} ja_j$,
%does not change: the change in the amount of cache used is given by 
%\begin{align}
%& \sum_{j=0}^{K} j\bar{a}_j - \sum_{j=0}^{K} ja_j \nonumber \\
%=&\left( i_1(a_{i_1}-\Delta) + (i_1+1)(a_{i_1+1}+\Delta) \right. \nonumber \\
%&\left. + (i_2-1)(a_{i_2-1}+\Delta) + i_2(a_{i_2}-\Delta) \right) \nonumber \\
%& - \left( i_1a_{i_1} + (i_1+1)a_{i_1+1} + (i_2-1)a_{i_2-1}+i_2a_{i_2} \right) \nonumber \\
%= &\, \Delta\left( -i_1 + (i_1+1) + (i_2 - 1) - i_2 \right) \nonumber \\
%= & \, 0. \nonumber 
%\end{align}
%
while the objective function decreases strictly, because the
coefficients of $a_j$ in the objective, $\frac{K-j}{j+1}$ is a
decreasing function of $j$ with decreasing second differences. 
%\begin{align}
%& \sum^{K}_{j=0} \frac{K-j}{j+1} \bar{a}_{j} - \sum^{K}_{j=0} \frac{K-j}{j+1} a_{j} \nonumber \\
%= & \, \Delta\left( \left(\frac{K-i_1-1}{i_1+2} - \frac{K-i_1}{i_1+1}\right)\right. \nonumber \\
%& - \left.\left(\frac{K-i_2}{i_2+1} - \frac{K-i_2+1}{i_2} \right)  \right), \label{doublediff} 
%\end{align}
%which is less than zero: both differences in \eqref{doublediff} are negative, because the objective function coefficients are decreasing functions of the index, and the first difference is larger in magnitude than the second difference, because the objective function coefficients themselves have positive second differences. Thus $\bar{\mathbf{a}}$ yields a better objective function value than $\mathbf{a}$, which concludes the proof.  

%Lemma \ref{prooflemma2} has some important consequences. Specifically, it 

The above two observations imply that 
%rule out any feasible solution that has three or more non-zero $a_j$ variables, because that would necessarily require at least two of the variables to have indices greater than two apart. This implies that 
the optimal solution has either only one
non-zero variable, or two non-zero variables that have adjacent
indices, i.e. some $j$ and $j+1$. %The proof of Theorem \ref{biggie} can now be stated.
%\emph{Proof of Theorem \ref{biggie}}: From Lemma \ref{prooflemma1}, the cache constraint \eqref{23}  must be satisfied with equality, and from Lemma \ref{prooflemma2}, there are at most two non-zero variables, which must have adjacent indices. Thus, 
In this case, the constraints for the optimal solution to \eqref{orig3}-\eqref{33} reduce to 
\begin{equation}
ja_j + (j+1)a_{j+1} = t, \label{cache}
\end{equation}
and 
\begin{equation}
a_j + a_{j+1} = 1. \label{simplex}
\end{equation}
%for an appropriate index $j$, and with $a_j \geq 0$ and $a_{j+1} \geq 0$. But the 
This system of equations %\eqref{cache}-\eqref{simplex} has a
has a unique solution.  Indeed, due to \eqref{simplex} and the positivity
constraints, \eqref{cache} implies that $t$ must be a \emph{convex
combination} of $j$ and $j+1$. %Thus it must be the case that $j \leq KM/N$, and $j+1 \geq KM/N$. 
If $t \notin \mathbb{Z}$, the only integer possibility for $j$ is $j =
\lfloor t \rfloor$ and $ j+1 = \lceil t \rceil$; representing $t=
s\lfloor t \rfloor + (1-s)\lceil t \rceil$ for some $s \in (0,1)$, it
becomes clear that the unique solution to
\eqref{cache}-\eqref{simplex} is $a_{j} = s, a_{j+1} = 1-s$ for 
$j = \lfloor t \rfloor$ and $s=\lceil t \rceil - t$.  
If $t \in \mathbb{Z}$, then we can set 
$j = t$, then the unique solution to \eqref{cache}-\eqref{simplex} is
$a_j = 1, a_{j+1} = 0$. Using the change of variables $v_i =
a_i/\binom{K}{i}$, this gives the optimal solution to
\eqref{orig2}-\eqref{32} as stated by \eqref{firstpart}-\eqref{secondpart}. 
% Using the aforementioned change of variables, this gives the optimal solution to \eqref{orig2}-\eqref{32} described by \eqref{firstpart}, which concludes the proof of the Theorem.
\end{IEEEproof}

As a final remark for this section, we acknowledge that there exist
stronger results about optimal coded caching schemes than Theorem
\ref{biggie} in the literature, for instance,
\cite{2016arXiv160907817Y} shows that a slightly modified version of
the scheme in \cite{6620392, 6763007} is the optimal coded caching
scheme among all schemes with uncoded cache content using information
theoretical upper bounds. %\footnote{The phrase ``coded caching'' is unfortunately used to refer both to schemes with and without coded cache content. The ``coded" in ``coded caching'' refers to the coded transmissions used.} 
The optimization theoretic perspective of this paper is nevertheless
worthwhile in that it easily begets extensions to more practical
non-uniform scenarios, which is the focus of the rest of this paper.

\section{Coded Caching with Heterogeneous Files}
\label{NU1}

We now move onto the coded caching problem with heterogeneous
parameters.  Although the optimization problem
\eqref{generalorig}-\eqref{generalconstraint} is already capable of
accounting for non-uniform file popularity, file length, and cache size,
the problem size scales exponentially in system parameters, hence the
optimization problem is intractable for practical system sizes. The main
contributions of this and the next section are to develop simplifications
to \eqref{generalorig}-\eqref{generalconstraint} that reduce the
computational complexity of the problem while maintaining high-quality
performance. Each non-uniformity is considered both individually and
in conjunction with the others in order to gain insight into the
interactions of their respective effects.  We begin by examining the
effect of non-uniform file popularity and non-uniform file length, but
for now keep the cache sizes uniform across the users.

%\section{Coded Caching with Hetereogeneous Files}
%%\label{ETMPC}

The procedure for each case is roughly as follows: first a new set 
of variables are defined that are intended to capture some 
structural feature of the problem, e.g. the $v_j$ variables of the 
simplified homogeneous problem \eqref{orig2}-\eqref{32}. Next,
certain conditions, referred to as \emph{memory inequality} constraints,
are imposed on the new variables (see e.g. \eqref{3rdfifthconstraint}), 
which forces feasible solutions to dedicate more cache memory to certain 
kinds of files, e.g. more popular files. While no \emph{a priori} justification 
for the use of these variable and constraints is given, subsequent numerical
results justify their use \emph{a posteriori}. 

These memory inequality constraints  
then allow the simplification of the objective function \eqref{generalorig}. 
First, the \emph{max} function can be eliminated from \eqref{generalorig}, since
the memory inequality constraints are sufficient to determine \emph{a priori} which subfiles 
will be the maximum given some request vector $\mathbf{d}$. 
This in turn allows the expected rate 
to computed precisely in terms of \emph{the largest file requested},
 \emph{the second largest file requested}, and so on, instead 
 of in terms of the $N^K$ possible request vectors. Since there
 are $K$ files requested, and $N$ possibilities for each file, this ultimately
 reduces the scaling of the objective function from exponential to polynomial, 
 although the objective function does not necessarily scale with $NK$ precisely.

%The rest of this paper is dedicated to developing high-quality but
%tractable simplifications of 
%\eqref{generalorig}-\eqref{generalconstraint} for cases with one
%or more sets of non-uniform systems parameters. 

\subsection{Prior Work}

The effect of non-uniform file popularity, also referred to as non-uniform demands, on coded caching has been explored in a number of papers \cite{6849235,7782760, 7904696, 6933485, 2016arXiv160505026C, 2016arXiv160905836H, 2016arXiv160905831H, 7218445,7865913, 6874794, 7282746,7308972}. In \cite{6849235,7782760}, Niesen and Maddah-Ali modify their decentralized scheme from \cite{6807823} by first grouping users according to the popularity of their respective file requests, and then transmitting to the groups sequentially using the scheme from \cite{6807823}. The authors of \cite{7904696, 6933485, 2016arXiv160905836H, 2016arXiv160905831H, 2016arXiv160505026C} develop an order-optimal scheme using random caching with a graph-based algorithm to design coded multicast transmissions, with \cite{2016arXiv160505026C} focusing specifically on the application of video delivery. In \cite{2016arXiv160905836H, 2016arXiv160905831H}, both the popularity of the files and their request correlation are considered when designing the caching and transmission scheme, while in \cite{7218445,7865913,6874794,7282746}, a heterogeneous network structure is considered, with file popularity organized in discrete levels. Using a novel random caching-based scheme, \cite{7308972} is able to show order-optimality with a constant that is independent of the popularity distribution. Finally, we repeat earlier comments that \cite{2017arXiv170707146J} uses the same optimization framework that is used to study non-uniform popularity in great depth; we nevertheless show our (similar) work for the sake of exposition. 
%% "Multi-level coded caching" appears to users centralized seen./designed caching, while their newer journal paper appears to use decent. caching/ random caching...

The literature on the effects of non-uniform file length is, however,
comparatively scarce. To the best of our knowledge, Zhang et al
\cite{7282743} provide the only scheme designed to accommodate
non-uniform file length for a general number of users. A scheme is provided that uses random caching with a transmission scheme similar to the one used in this paper, and upper and lower bounds on system performance are derived. In particular, \cite{7282743}  explores a random caching scheme where files are cached with a probability proportional to size of the file. Non-uniform file size is also explored in the recent letter \cite{7523930}, in which the achievable rate region for both non-uniform file size and non-uniform cache size is characterized, but only for the case of $K=2$ users and $N=2$ files. 

Note that it can be argued that if files are indeed different sizes,
they can be broken up into smaller packets of a constant size $F'$
bits, and then treated as separate files. While this is a reasonable
assumption while investigating other aspects of a coded caching
scheme, there are two issues that need be addressed in practice.
First, if a file is broken up into multiple pieces, then a user who
seeks to to download the entire file must make multiple (correlated)
requests to the server - a fact that should be accounted for in
subsequent system design. The second practical issue comes from the
fact that it is unclear how to set the common file size $F'$:
efficiency demands that $F'$ be as large as possible so that any
required headers represent a small proportion of the entire download,
while at the same time, $F'$ should also be small enough to divide
files without significant remainder.

We therefore contend that heterogeneous file length is an important
parameter that a practical system must capable of accommodating in one
way or another.  The approach to handling non-uniform files sizes
discussed above may indeed have some merit; some work has been done to
analyze the case of multiple requests from users, see e.g.
\cite{7282744,7843674,2015arXiv151107542J,2014arXiv1402.4572J}, and so
an approach based on this technique may be viable. This paper,
however, uses a different approach: files are not broken up into
smaller files of equal lengths, and so the cache content is designed to
accommodate their different lengths. The possibility of comparing the
performance of the different approaches is left to future work.

To the best of our knowledge, there has been no work exploring the
relationship between file length and popularity and the resulting effect on cache content. In the literature discussed above, it is noted (roughly) that more popular files should be allocated more cache memory, but also that larger files should be allocated more cache memory. Given that, in general, file lengths and popularity may not have any correlation, it is not clear how these non-uniformities jointly affect optimal cache content.  These interactions are explored in the following.

\subsection{Preliminaries} 

First, we
introduce two lemmas. The first one is a classic result about binomial coefficients, %the proof of which can readily be found online by the interested reader,
while the second lemma is used to determine the probability that the file $n$ is the $k$-th largest file requested; the proof of the latter is contained in Appendix \ref{B}.

\begin{lemma}[Chu-Vandermonde Convolution]
\label{CVC}
For $N,N_1,N_2,$ and $n$ positive integers, with $N_1 + N_2 = N$ and $n \leq N$, 
\begin{equation}
\binom{N}{n} = \sum_{k = 0}^{n}\binom{N_1}{k}\binom{N_2}{n-k} \label{CVCeq}
\end{equation}
\end{lemma}

\begin{lemma}
\label{problemma}
Consider $K$ independent multinomial random trials %vector $\mathbf{X} \in [K]^N $ 
with $N$ possible outcomes per trial, with probabilities $\{p_1, p_2, \dots, p_N\}$,
denoted by $\mathbf{Z} \in [N]^K$, i.e., $Z_i$, 
the $i$-th element of $\mathbf{Z}$, 
is the outcome of the $i$-th trial. %, and the $n$-th element of $\mathbf{X}$, $X_n$, counts the number of times outcome $n$ is drawn in the $K$ trials, i.e. the number of times outcome $n$ appears in $\mathbf{Z}$. 
Let the random vector ${\bf Y}$ be a sorted version of ${\bf Z}$, but with 
index shifted by 1, 
%denote the ($m+1$)-th smallest element of $\mathbf{Z}$; so that $Y_0$
%Let the random variable $Y_m$ denote the ($m+1$)-th smallest element of $\mathbf{Z}$; 
so that $Y_0$ is the smallest element of $\mathbf{Z}$, $Y_1$ is the second smallest element, and so on. 
The probability mass function of $Y_m$ is given by 
\begin{flalign}
\text{Pr}[Y_0 = i ] = &\,\, \left(\sum_{l=i}^N p_l\right)^K - \left(\sum_{i+1}^N p_l\right)^K \label{m0case}, \\
\text{Pr}[Y_1 = i ] = &\,\, \text{Pr}[Y_0 = i] + K\left(\left(\sum_{l=1}^{i-1}p_l\right)\left(\sum_{l=i}^{N}p_l\right)^{K-1} \right. \nonumber \\ 
&-\, \left. \left(\sum_{l=1}^{i}p_l\right)\left(\sum_{l=i+1}^{N}p_l\right)^{K-1}\right), \label{m1case} 
\end{flalign}
and for $m \in [2:K-1] $, 
\begin{flalign}
\text{Pr}[ Y_m = 1] =& \sum_{k=0}^{K-m-1}\binom{K}{m+1+k} \nonumber \\
& \qquad \qquad p_1^{m+1+k}(1-p_1)^{K-m-1-k} 
\end{flalign}
\begin{multline}
\text{Pr}[Y_m = i] = \\
\binom{K}{K-m}\left(\left(\sum_{l=i}^{N}p_l\right)^{K-m} -
\left(\sum_{l=i+1}^{N}p_l\right)^{K-m} \right) \\
\left(\sum_{l=1}^{i-1}p_l\right)^m + \sum_{k=0}^{K-2}\sum_{b =
\max{\{0,m-1-k\}}}^{\min{\{m-1,K-2-k\}}} \\
\left( \vphantom{\left(\sum_{l=1}^{i-1}p_l\right)^b} 
\binom{K}{2+k,b,K-2-k-b}
p_i^{2+k} \right. \\
\left.
\left(\sum_{l=1}^{i-1}p_l\right)^b\left(\sum_{l=i+1}^{N}p_l\right)^{K-2-k-b} \right) \label{melseelsecase}
\end{multline}
where the final expression is for $i \in [2:N]$. %\footnote{We remind the reader that by the notation used in this paper, $\sum_{l=a}^b n_l = 0$ and $(\sum_{l=a}^b n_l)^0 = 1 $ for $a > b$.} 
\end{lemma}

\subsection{Optimization Formulation}
 \label{PF1}

The first step in reducing the exponential number of variables in
\eqref{generalorig}-\eqref{generalconstraint} is 
the definition of a new set of $(K+1)N$ variables as
\begin{equation}
 v_{l,j} = |W^{(l)}_{\mathcal{S}}|, \forall \mathcal{S} \text{ s.t. } |\mathcal{S}| = j, \forall l \in [N]. \label{firstnewvar}
 \end{equation}
similar to the $v_j$ variables used in Section \ref{AOTPOFLC}, except
now there is a set of $v_j$ variables for each file to capture
the difference in length and popularity between files.  
Such a reduction enforces $|W^{(l)}_{\mathcal{S}}|$ to
depend only on the cardinality of $\mathcal{S}$, which is a reasonable
thing to do and in fact can be proved to be without loss of generality
for the special case of non-uniform file popularity alone but with uniform file
length \cite{2017arXiv170707146J}.
The effect of this reduction in the general case
is numerically evaluated later in the section. 

%These $(K+1)N$ variables may result in a loss of generality, but are
%useful for reducing the complexity of the problem
%\eqref{generalorig}-\eqref{generalconstraint}. Note that these are

The simplification of the general constraints in
\eqref{generalconstraint} follows similar reasoning used to
obtain the constraints \eqref{2ndfirstconstraint}-\eqref{2ndsecondconstraint} in Section \ref{AOTPOFLC}, except now there are arbitrary file lengths $F_l$ and popularities $p_l$; this gives
 \begin{equation}
 \sum_{j=0}^{K}\binom{K}{j}v_{l,j} = F_l, \forall l \in [N] \label{3rdfirstconstraint} 
 \end{equation}
as the file reconstruction constraints, and 
\begin{equation}
\sum_{j=1}^{K} \binom{K-1}{j-1}v_{l,j} \leq \mu_l, \forall{l} \in [N] \label{3rdsecondconstraint}
\end{equation}
for the cache constraint. The other two constraints,
\begin{equation}
v_{l,j} \geq 0, \forall l \in [N], \forall j \in [0:K], \label{3rdthirdconstraint}
\end{equation}
and
\begin{equation}
\sum_{l = 1}^N \mu_l \leq M \label{3rdfourthconstraint}
\end{equation}
have more obvious modifications. 

To express the objective function in polynomial number of terms, we
now need to impose certain \emph{memory inequality} conditions in
order to simplify the max operator in the objective. We propose
two different approaches called the \emph{popularity-first} approach
and the \emph{length-first} approach respectively for handling the
non-uniform file popularity and file length. 

\subsubsection{Popularity-First Approach} 

In the popularity-first approach, files are labelled in decreasing
over of popularity, i.e. such that $p_1 \geq \dots \geq p_N$. Then,
motivated by the idea that more popular files ought to have more cache
space dedicated to them, a memory inequality condition is imposed on
the cache content as 
\begin{equation}
v_{l_1,j} \geq v_{l_2,j} ,\forall l_1,l_2 \in [N] \text{ s.t. } l_1 < l_2, j \in [K].\label{3rdfifthconstraint}
\end{equation}
This memory inequality constraint is adopted to help reduce the
complexity of the problem; as previously discussed, this constraint
(and others like it later in the paper) allow the \emph{max} function
in \eqref{generalorig} to be eliminated in favour of a linear function
of the variables, which in turn allows for the expected rate to be
computed in a polynomial number of operations, instead of the
exponential number required by \eqref{generalorig}. 

The popularity-first approach is most appropriate for the special case 
of non-uniform file popularity alone, but with uniform file length. 
The following proposition shows explicitly the effect that \eqref{3rdfifthconstraint} has on the objective function in this case.
Note that in this special case of uniform file length,
\eqref{3rdfifthconstraint}, which holds for $j= 1, \dots, K$, becomes reversed for $j=0$. 
To see this, consider two files $l_1$ and $l_2$ with $l_1 < l_2$, that satisfy \eqref{3rdfifthconstraint}; if both have length $F$, i.e. satisfy \eqref{3rdfirstconstraint} with $F_{l_!} = F_{l_2} = F$, then $v_{l_1,0} \leq v_{l_2,0}$ because every other subfile of $l_1$ is larger than every other subfile of $l_2$.

As mentioned earlier, this special case of non-uniform file popularity
alone with uniform file length has already been considered in
independent work \cite{2017arXiv170707146J}. But the problem
formulation of \cite{2017arXiv170707146J} does not account for the difference
in the lengths of subfiles within the max operator, thus may result
in loss of optimality. The expression below is an exact accounting of
the expected delivery rate.

\begin{proposition}
 \label{NUPobjfunc}
Consider the case of non-uniform file popularity and uniform file
length.  Let the variables defined in \eqref{firstnewvar} be subject
to condition \eqref{3rdfifthconstraint} with files labelled in
decreasing order of popularity. Then the objective function
\eqref{generalorig} simplifies exactly as 
\begin{multline}
\mathbb{E}\left[\sum_{\mathcal{S}\in \mathcal{P}(\mathcal{U})\setminus \emptyset}\max_{k \in \mathcal{S}}\{|W_{S\setminus\{k\}}^{(d_k)}|\} \right] = \\
\sum_{j=1}^{K-1} \sum_{i=0}^{K-1} \sum_{l=1}^{N} 
\binom{K-1-i}{j}\text{Pr}[Y_i=l]v_{l,j} \\
  +\sum_{i=0}^{K-1} \sum_{l=1}^{N} \text{Pr}[Y_{K-i-1} = l]v_{l,0}, \label{simpobjfunc}
 \end{multline}
 where $Y_i$ is the random variable representing the ($i+1$)-th smallest index in a random request vector $\mathbf{d}$.%, while $\tilde{Y}_i$ is the random variable representing the ($i+1$)-th largest index in a random request vector $\mathbf{d}$. 
 \end{proposition} 
 
 The proof of Proposition \ref{NUPobjfunc} is contained in Appendix
\ref{C}. Note that the probabilities $\text{Pr}[Y_i = l]$ can be
obtained directly from Lemma \ref{problemma}: since the files are labelled in decreasing order of popularity, the probability that $l$ is the $(i+1)$-th smallest file requested in $\mathbf{d}$ is equivalent to the probability that $l$ is the $(i+1)$-th smallest index in $\mathbf{Z}$.
%, while the
%$\text{Pr}[\tilde{Y}_i = l]$ probabilities can be obtained from Lemma
%\ref{problemma} by reversing the labels of the files, i.e. such that
%the smallest index in $\mathbf{d}$ is now the largest, and so on, due
%to that the memory inequality for $j=0$ is reversed. 
The optimization
problem for the case of non-uniform file popularity, uniform file
length, and uniform cache size can now be written as 
\begin{align}
\mbox{minimize} \;\;\; & \sum_{j=1}^{K-1} \sum_{i=0}^{K-1} \sum_{l=1}^{N} \binom{K-1-i}{j}\text{Pr}[Y_i=l]v_{l,j}\nonumber \\
 & \quad +\sum_{i=0}^{K-1} \sum_{l=1}^{N} \text{Pr}[Y_{K-i-1} = l]v_{l,0}   \label{3rdobjfunc}\\
\mbox{subject to} \;\;\; & \text{ \eqref{3rdfirstconstraint}-\eqref{3rdfifthconstraint},} \label{3rdconstraints}
\end{align}
Note that this is a linear program, with a number of summands in the
objective function that scales $K^2N$, and exactly $KN$ variables and
$N(K+3)+K(N(N-1))/2+1$ constraints. %Note also that this is similar to ``Problem 3'' in \cite{2017arXiv170707146J}.

If the files are also of non-uniform length, %a slight modification is required. Letting $f(i)$ denote the index of the $i$-th biggest file in the set of files, the memory inequality constraints used for this case are
%\begin{equation}
%v_{l_1,j} \geq v_{l_2,j} ,\forall l_1,l_2 \in [N] \text{ s.t. } l_1 < l_2, j \in [K-1] ,\label{5thfifthconstraintp1}
%\end{equation}
%and
%\begin{equation}
%v_{f(i_1),0} \geq v_{f(i_2),0} ,\forall i_1,i_2 \in [N] \text{ s.t. } i_1 < i_2.\label{5thfifthconstraintp2}
%\end{equation}
%Essentially, files are ordered by probability for $j > 0$ and by size for $j=0$. Using the same reasoning as in Proposition \ref{NUPobjfunc}, the simplified objective function in this case is
%\begin{align}
%&\sum_{j=1}^{K-1} \sum_{i=0}^{K-1} \sum_{l=1}^{N} \binom{K-1-i}{j}\text{Pr}[Y_i=l]v_{l,j}\nonumber \\
% & +\sum_{i=0}^{K-1} \sum_{l=1}^{N} \text{Pr}[\hat{Y}_{i} = f(l)]v_{f(l),0}, \label{OtherNUPof}
%\end{align}
%where $\hat{Y}_i$ is the random variable representing the index of $(i+1)$-th biggest file in the request vector $\mathbf{d}$. The probabilities Pr[$\hat{Y}_i = f(l)$] can be computed using Lemma \ref{problemma} by ordering the files in terms of decreasing length. The popularity-first optimization problem for non-uniform file popularity, non-uniform file length, and uniform cache size is thus
%\begin{align}
%\mbox{minimize} \;\;\; &\sum_{j=1}^{K-1} \sum_{i=0}^{K-1} \sum_{l=1}^{N} \binom{K-1-i}{j}\text{Pr}[Y_i=l]v_{l,j}\nonumber \\
% & +\sum_{i=0}^{K-1} \sum_{l=1}^{N} \text{Pr}[\hat{Y}_{i} = f(l)]v_{f(l),0}, \label{5thobjfunc}\\
%\mbox{subject to} \;\;\; & \text{ \eqref{3rdfirstconstraint}-\eqref{3rdfourthconstraint}, \eqref{5thfifthconstraintp1}, \eqref{5thfifthconstraintp2}.} \label{5thconstraints}
%\end{align}
further work is required, but a similar optimization problem can
nonetheless be developed. The details are omitted here both for the
sake of brevity and because numerical results suggest that the
popularity-first approach does not perform as well as the length-first
approach in the general case when both popularity and file lengths are
non-uniform. %in the general case, at least for the numerical evaluations considered in this paper (see Figure \ref{NUPFpic}).  

\subsubsection{Length-First Approach} 

In the length-first approach, files are labelled in decreasing order of length, i.e. such that $F_1 \geq F_2 \geq \dots \geq F_N$. Then, motivated by the idea that longer files ought to have more cache space dedicated to them, the following memory inequality condition is imposed:
\begin{equation}
v_{l_1,j} \geq v_{l_2,j} ,\forall l_1,l_2 \in [N] \text{ s.t. } l_1 < l_2, j \in [0:K-1]. \label{4thfifthconstraint}
\end{equation}

Using similar reasoning as Proposition \ref{NUPobjfunc}, it is straightforward to show 
that
\begin{proposition}
 \label{NUPobjfunc2}
Consider the case of non-uniform file length with either uniform or
non-uniform popularity.  Let the variables defined in \eqref{firstnewvar} be subject
to condition \eqref{4thfifthconstraint} with files labelled in
decreasing order of length. Then the objective function
\eqref{generalorig} simplifies exactly as 
\begin{multline}
\mathbb{E}\left[\sum_{\mathcal{S}\in \mathcal{P}(\mathcal{U})\setminus
\emptyset}\max_{k \in \mathcal{S}}\{|W_{S\setminus\{k\}}^{(d_k)}|\} \right] \\
 = \sum_{j=0}^{K-1} \sum_{i=0}^{K-1} \sum_{l=1}^{N} \binom{K-1-i}{j}\text{Pr}[Y_i=l]v_{l,j} \label{simpobjfunc2}
 \end{multline}
where $Y_i$ is the random variable representing the ($i+1$)-th
smallest index in a random request vector $\mathbf{d}$. 
 \end{proposition} 
 
The length-first optimization problem for non-uniform file popularity, non-uniform file length, and uniform cache size is obtained as
\begin{align}
\mbox{minimize} \;\;\; & \sum_{j=0}^{K-1} \sum_{i=0}^{K-1} \sum_{l=1}^{N} \binom{K-1-i}{j}\text{Pr}[Y_i=l]v_{l,j}\label{4thobjfunc} \\
\mbox{subject to} \;\;\; & \text{ \eqref{3rdfirstconstraint}-\eqref{3rdfourthconstraint}, \eqref{4thfifthconstraint}} \label{4thconstraints}
\end{align}
which is a linear program with $K^2N$ summands in the objective
function, $KN$ variables, and $N(K+3)+K(N(N-1))/2 + 1$ constraints. 
%Note that this problem can be used for both uniform and non-uniform file popularities. 

\subsection{Numerical Results}

To evaluation the effect of the simplified problem formulation, we
consider a case with $K=4$ users with equal cache sizes of $M$, and
$N=6$ files. When the files are of uniform length, the value $F = 1$
is used, and when they are of non-uniform length, the values $\{F_1,
\dots, F_6\} = \{9/6, 8/6, 7/6, 5/6, 4/6, 3/6\}$ are used. Similarly,
when the files are of uniform popularity, the value $p_l = 1/N$ is
used for all $l \in [N]$, but when the files have non-uniform
popularity, the distribution is given by a Zipf distribution with
parameter $s$, which has been observed empirically to be reasonable
model for user demands; a parameter of $s = 0.56$ is used in this
paper (see e.g. \cite{6600983}). When both the file lengths and
popularities are non-uniform, the relationship between length and
popularity is specified explicitly. %Both problems (and indeed, every subsequent optimization problem in this paper) are solved using the convex optimization solver CVX \cite{cvx,gb08}. 
 
\begin{figure}%[htbp!]
\centering
\includegraphics[width=3.6in]{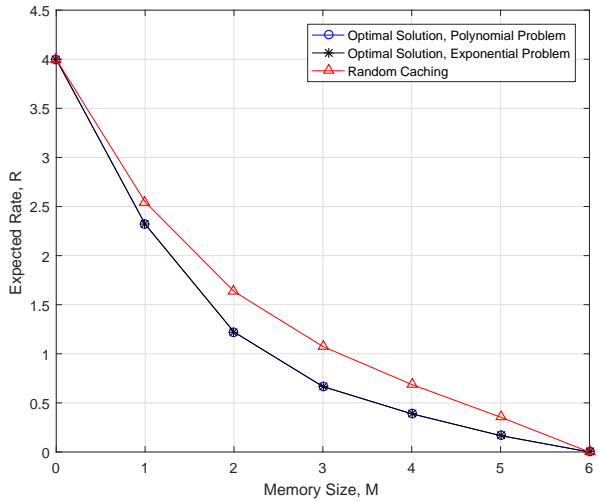}
\caption{\label{NUPpic} A comparison of the performance of the solution obtained from \eqref{generalorig}-\eqref{generalconstraint} to the solution obtained by \eqref{3rdobjfunc}-\eqref{3rdconstraints}, with reference to a baseline random caching scheme, for the case of non-uniform file popularity only}
\end{figure}

Fig.~\ref{NUPpic} compares the rate-memory tradeoff curve for the
original problem \eqref{generalorig}-\eqref{generalconstraint} and the
simplified problem \eqref{3rdobjfunc}-\eqref{3rdconstraints} for the
non-uniform popularity and uniform length case. A baseline random caching scheme is also included for reference. This random caching scheme is essentially the decentralized scheme of \cite{6807823} but with file $n$ allocated $\mu_nF$ bits of cache memory instead of $MF/N$ bits; initially, the value is obtained as $\mu_n = \min\{Mp_n,1\}$ for all $n \in [N]$, and if $\sum_{n=1}^N \mu_n < M$ after that, the remaining cache memory is allocated to each file sequentially until the remaining memory runs out. 

\begin{figure}%[htbp!]
\centering
\includegraphics[width=3.6in]{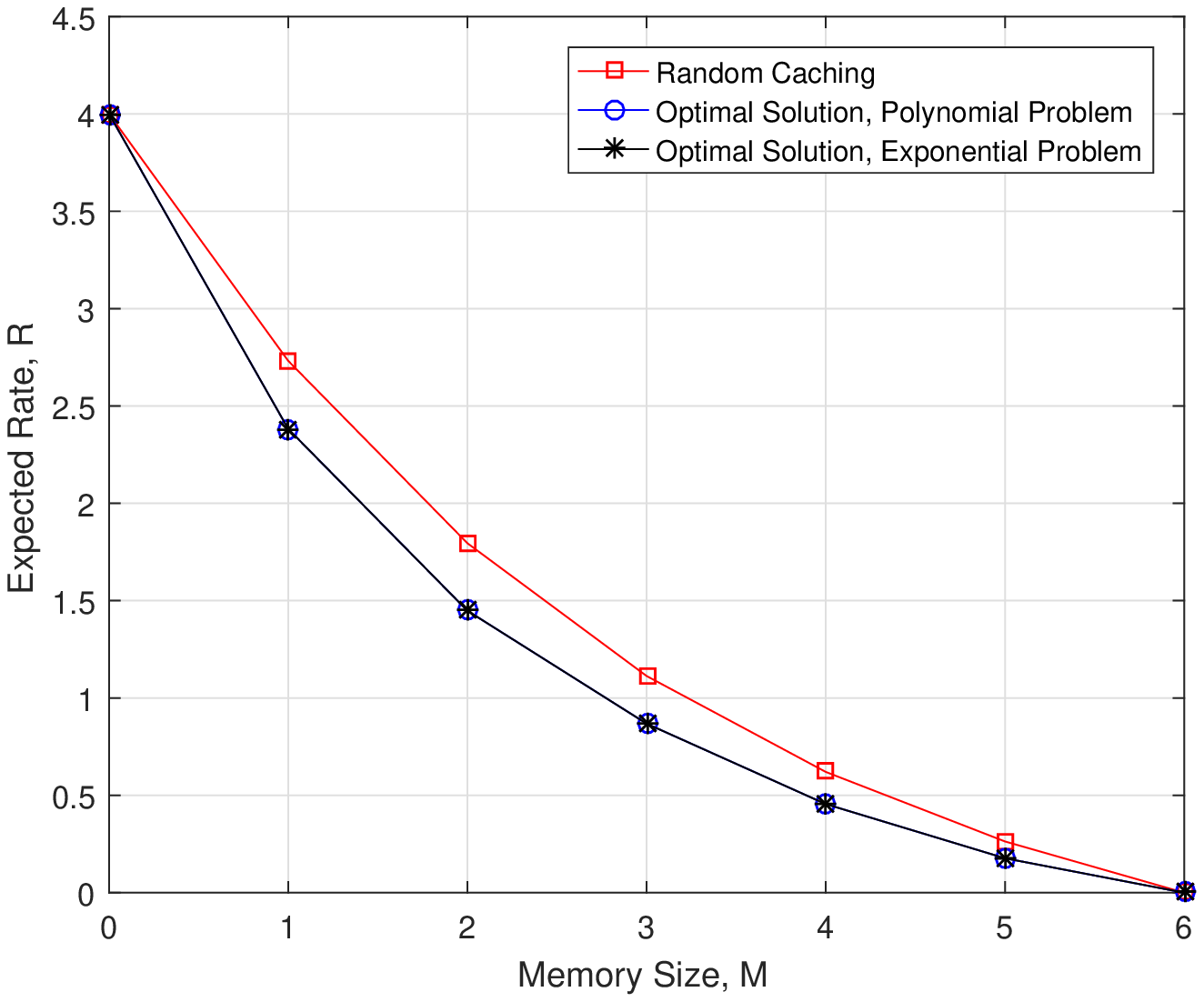}
\caption{\label{NUFpic} A comparison of the performance of the solution obtained from \eqref{generalorig}-\eqref{generalconstraint} to the solution obtained by \eqref{4thobjfunc}-\eqref{4thconstraints}, with reference to a baseline random caching scheme, for the case of non-uniform file length only.}
\end{figure}

Conversely, Fig.~\ref{NUFpic} compares the rate-memory tradeoff curves
of \eqref{generalorig}-\eqref{generalconstraint} and the simplified
problem \eqref{4thobjfunc}-\eqref{4thconstraints} for the non-uniform
length and uniform popularity case. The random caching baseline scheme
used here is essentially equivalent to the one proposed in
\cite{7282743}.

Both figures show that the performance of the general problem and the two simplified problems is identical for the respective cases considered here. Indeed, when the numerical solutions of \eqref{generalorig}-\eqref{generalconstraint} for these two cases are examined explicitly, it is clear that the memory constraint conditions are indeed satisfied, and so the optimal solutions in these cases are attainable by the respective simplified problems.

\begin{figure}%[htbp!]
\centering
\includegraphics[width=3.6in]{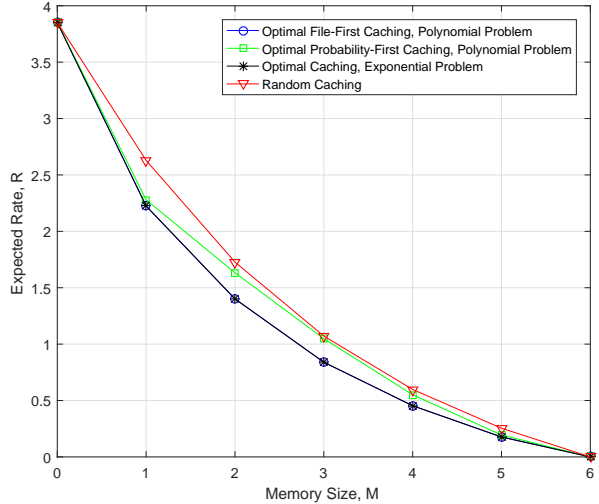}
\caption{\label{NUPFpic} A comparison of the performance of the solution obtained from \eqref{generalorig}-\eqref{generalconstraint} to the solutions obtained by \eqref{4thobjfunc}-\eqref{4thconstraints} and %\eqref{5thobjfunc}-\eqref{5thconstraints}
a popularity-first optimization problem, with reference to a baseline random caching scheme, for the case of non-uniform file length and popularity.}
\end{figure}

Next, Fig.~\ref{NUPFpic} compares the performance of the original
problem \eqref{generalorig}-\eqref{generalconstraint} to both the
length-first and popularity-first simplified problems for the case of non-uniform file length and popularity. The specific pairings of length and popularities used and their associated labels are listed in Table \ref{NUPFOrder1}. %and \ref{NUPFOrder2}, which show the labelling used for the length-first and probability-first problems respectively. 

Although somewhat arbitrary, these file length and popularity combinations are intended to simulate a practical scenario where file popularity and length are relatively uncorrelated. While examining this individual case is not sufficient for determining general patterns, it is enough to gain some important insight about the tension between file length and popularity. 

%\begin{table}
%\centering
%\begin{tabular}{| c | c | c |}
%\hline
%\multicolumn{3}{|c|}{Popularity-First Labelling} \\
%\hline
%File Index & Length & Popularity \\
%\hline
%1&4/6& 0.2897 \\
%\hline
%2&7/6 &0.1965  \\
%\hline
 %3&8/6&  0.1566\\
%\hline
 %4&3/6 &  0.1333 \\
%\hline
 %5&9/6 &  0.1176\\
%\hline
 %6&5/6 &  0.1062\\
%\hline
%\end{tabular}
%\caption{The file labelling used to compute the caching scheme for problem \eqref{5thobjfunc}-\eqref{5thconstraints} in the non-uniform file popularity and length case.}
%\label{NUPFOrder2}
%\end{table}

\begin{table}
\centering
\caption{The file labelling and length-popularity pairings used in the non-uniform file popularity and length case. }
\label{NUPFOrder1}
\begin{tabular}{| c | c | c | c |}
\hline
\multicolumn{4}{|c|}{Length-First (LF) and Popularity-First (PF) Labelling} \\
\hline
LF File Index &PF File Index& Length & Popularity \\
\hline
1&5&9/6& 0.1176 \\
\hline
2&3&8/6 & 0.1566 \\
\hline
 3&2&7/6&  0.1965\\
\hline
 4&6&5/6 & 0.1062  \\
\hline
 5&1&4/6 &  0.2897\\
\hline
 6&4&3/6 &  0.1333\\
\hline
\end{tabular}
\end{table}

Fig.~\ref{NUPFpic} shows that in this case, the length-first scheme yields much better results than the popularity-first scheme. Indeed, the length-first scheme obtains the same performance as the original problem for all $M$ considered except $M = 1$. The reason for this divergence can be seen from Table \ref{NUPFTable} which shows the optimal solution to the original problem \eqref{generalorig}-\eqref{generalconstraint} in the $M=1$ case. The value of $|W^{(l)}_{\mathcal{S}}|$ is shown in the $l$-th column of the row labelled with $\mathcal{S}$, and the files are ordered using the length-first labelling of Table \ref{NUPFOrder1}. It is clear that file 1, the largest file but fifth-most popular, has been allocated less cache memory than files 2 and 3, which are the third and second most popular files respectively. Thus the memory-inequality constraint \eqref{4thfifthconstraint} is violated and so the length-first simplified problem cannot attain the optimal solution of \eqref{generalorig}-\eqref{generalconstraint}. Despite this, the optimal value to the simplified problem is only about $10^{-4}$ larger than the optimal value of \eqref{generalorig}-\eqref{generalconstraint}, and so the difference is not significant. 
It would thus appear that, while probability cannot be completely ignored in theory, a length-first approach to caching can yield very good results in practice. This insight is later used in Section \ref{finprobform}, but first the problem of non-uniform cache size must be studied on its own first; this is done next.

\begin{table}
\centering
\caption{Optimal subfile sizes and memory allocation for the general problem \eqref{generalorig}-\eqref{generalconstraint} with $K=4, N = 6$, $M = 1$, and file lengths and popularities given in Table \ref{NUPFOrder1} (using length-first indexing), with values rounded to three decimal places.}
\label{NUPFTable}
\begin{tabular}{|c|c|c|c|c|c|c|}
\hline
& \multicolumn{6}{|c|}{File Index}\\
\hline
Subset &1&2&3&4&5&6  \\ 
\hline
$\emptyset$ & 0.833 & 0.583& 0.417 & 0.167& 0& 0 \\ 
\hline
$\{1\} $& 0.167 & 0.188& 0.188 &0.167 &0.167&0.125 \\ 
\hline
$\{2\}$ & 0.167 & 0.188& 0.188 &0.167 &0.167&0.125 \\ 
\hline
$\{3\}$ & 0.167 & 0.188& 0.188 &0.167 &0.167&0.125 \\ 
\hline
$\{4\} $& 0.167 & 0.188& 0.188 &0.167 &0.167&0.125 \\ 
\hline
$\{1,2 \} $&0  &0 &0 &0 & 0&0  \\ 
\hline
 $\dots$ & \multicolumn{6}{|c|}{$\dots$}\\
\hline
$\{1,2,3,4\} $& 0 &0&0 & 0&0& 0 \\
\hline 
Total memory: & 0.167 & 0.188 & 0.188& 0.167 & 0.167 & 0.125 \\
\hline
\end{tabular}
\end{table}

\section{Coded Caching with Heterogeneous Cache Sizes}
\label{NUM}

%\subsection{Background}

We next consider simplifying the optimization formulation for the case with
non-uniform cache sizes. For now, file popularity and file length are
kept uniform; the case with all parameters being non-uniform is
treated in the subsequent section. For the case of non-uniform cache size,
%We move away from non-uniform file size and popularity for the moment to consider non-uniform cache size. 
a decentralized coded caching scheme is developed in
\cite{2015arXiv150401123W} and subsequently improved upon in the 
$K>N$ case by \cite{2016arXiv161101579A,7869142}. As previously
discussed, \cite{7925535} uses an optimization framework similar to
the one used in this paper to generate a scheme for the centralized
case. For the sake of completion, we note again the work
\cite{7523930} in which the rate region for both non-uniform cache 
and file size is characterized for $K=2$ users and $N=2$ files, but
the optimal scheme is not yet known for the general case.

\subsection{Optimization Formulation}
\label{NUMPF}

We first consider a simple case where there are only two cache
sizes, ``large'' and ``small'', represented by $M_L$ and $M_S$
respectively. The variable $K_L$ is used to represent the number of
users with large caches, and $K_S$ is used to represent the number of
users with small caches, such that $K_L + K_S = K$.
Note that file lengths and popularities are fixed as uniform, i.e. $p_1 = \dots = p_N
= 1/N$ and $F_1 = \dots = F_N = F = 1$ respectively. 
%, while considering non-uniform cache sizes at the user side.

Since files here are equally popular and of the same size, we have
symmetry across files, but now the symmetry across users is broken by
the non-uniform cache size. However, certain user symmetry still
exists, i.e., symmetry among members of the same cache size group. We therefore define three sets of variables for $j \in [0:K]$, denoted $v_{j,S}, v_{j,L}$, and $v_{j,M}$, as follows. For a subset of users $\mathcal{S}$ with a size $|\mathcal{S}| = j$, 
\begin{equation}
|W^{(l)}_{S}| = 
\begin{cases}
v_{j,S} & \text{if $\mathcal{S}$ contains only small-cache users,}  \\
v_{j,L} & \text{if $\mathcal{S}$ contains only large-cache users,} \\
v_{j,M} & \text{otherwise},  \\
\end{cases}
 \label{secondnewvar}
\end{equation}
for all files $l \in [N]$. 

This definition suggests some natural constraints. First, since there are only $K_S$ small-cache users, there cannot be a group of $j$ small-cache users if $j > K_S$, so we set
\begin{equation}
v_{j,S} = 0, \forall j > K_S. \label{firstofmany}
\end{equation}
Similarly, for large-cache users,
\begin{equation}
v_{j,L} = 0, \forall j > K_L. \label{secondofmany}
\end{equation}

A similar constraint is required for the $v_{1,M}$. Since there cannot be a subset of size one with both large- and small-cache users, we require 
\begin{equation}
v_{1,M} = 0; \label{thirdofmany}
\end{equation}
Moreover, since the $j=0$ variables correspond to subsets of size 0, it it not particularly meaningful to discuss whether or note this corresponds to a small, large, or mixed subset. To avoid any further complications, we simply set
\begin{equation}
v_{0,S} = v_{0,L} = v_{0,M} = v_0\, , \label{allequal}
\end{equation}

To understand the file reconstruction condition, note that, for $j \geq 2$, there are $\binom{K_S}{j}$ groups of small-cache users, $\binom{K_L}{j}$ groups of large-cache users, and 
\begin{equation}
\sum_{i = 1}^{j-1} \binom{K_S}{i}\binom{K_L}{j-i} \nonumber
\end{equation}  
mixed groups, because each mixed group must have at least one small-cache users and at least one-large cache user. By adding and subtracting the $i = 0$ and $i = j$ terms and using Lemma \ref{CVC}, we can rewrite this as 
\begin{multline}
\sum_{i = 0}^j \binom{K_S}{i}\binom{K_L}{j-i} - \binom{K_L}{j} - \binom{K_S}{j} \\
 = \binom{K}{j} - \binom{K_L}{j} - \binom{K_S}{j}, \nonumber
\end{multline}
and so the total portion of the file cached by subsets of size $j \geq 2$ is 
\begin{equation}
\binom{K_L}{j}(v_{j,L}-v_{j,M}) + \binom{K_S}{j}(v_{j,S}-v_{j,M}) + \binom{K}{j}v_{j,M}. \label{gettinthere}
\end{equation}
For $j = 1$, there are $K_L = \binom{K_L}{1}$ large users and $K_S = \binom{K_S}{1}$ small users. Moreover, since $v_{1,M} = 0$, it is easy to see that \eqref{gettinthere} also holds for $j = 1$. Finally, consider the $j = 0$ case. Here we only need to add $v_0$ to capture the portion of the file not cached by any user. However, note that if we set $j = 0$ in \eqref{gettinthere} and apply constraint \eqref{allequal}, the first two terms become 0, while the last term reduces to $v_{0,M} = v_{0}$. Thus \eqref{gettinthere} applies for all $j$, and so we can conveniently express the file reconstruction constraint as 
\begin{multline}
\sum_{j=0}^K \binom{K_L}{j}(v_{j,L}-v_{j,M}) + \binom{K_S}{j}(v_{j,S}-v_{j,M}) 
\\
+ \binom{K}{j}v_{j,M}  = 1. \label{6thfirstconstraint}
\end{multline}

Similar reasoning is used to obtain the cache memory constraints. A small cache user caches every subfile labelled with a subset in which he is contained as a member, and so necessarily caches only subfiles of size $v_{j,S}$ and $v_{j,M}$. Specifically, a small-cache user is a member of $\binom{K_S-1}{j-1}$ small groups of size $j$, and 
\begin{equation}
\sum_{i = 0}^{K-2} \binom{K_S-1}{i}\binom{K_L}{j-1-i}
\end{equation}
mixed groups. Using Lemma \ref{CVC} once again, the cache memory constraint for the small user is obtained as
\begin{align}
\sum_{j=1}^K  \binom{K_S-1}{j-1}(v_{j,S}-v_{j,M}) +\binom{K-1}{j-1}v_{j,M} = M_S/N. \label{6thsecondconstraintp1}
\end{align}
Note that the $j=1$ expression reduces to $\binom{K_L-1}{0}(v_{1,S}-0)+0= v_{1,S}$, as desired. Note also that each file is allocated an equal $M_S/N$ of the cache because the files are equally sized and equally popular. Using similar reasoning for large-cache users, we obtain
\begin{align}
\sum_{j=1}^K  \binom{K_L-1}{j-1}(v_{j,L}-v_{j,M}) 
+ \binom{K-1}{j-1}v_{j,M} = M_L/N. \label{6thsecondconstraintp2}
\end{align}

As always, it is required that all subfiles be of nonnegative size:
\begin{eqnarray}
v_{j,S} \geq 0, v_{j,L} \geq 0, v_{j,M} \geq 0, \forall j \in \{0, \dots, K\}. \label{6ththirdconstraint}
\end{eqnarray}
%We note that this introduces some redundancy into the constraints because some of these variables are already defined to be zero.

Finally, the memory inequality constraints for this problem are introduced. In general, we would expect the subfiles that large users have cached to be longer than the ones cached by small users. This is codified in the problem explicitly with 
\begin{flalign}
v_{j,L} \geq& \,\,v_{j,M}, \,\,j \in [2:K_L] \label{6thfifthconstraintp1} \\
v_{j,M}\geq&\,\, v_{j,S}, \,\,j \in [2:K_S] \label{6thfifthconstraintp2} \\
v_{j,L} \geq&\,\, v_{j,S}, \,\,j \in [1:K_L] \label{6thfifthconstraintp3}. 
\end{flalign}
Again note that there is some redundancy in these constraints, but they are nonetheless included for clarity of exposition. Note also that the first two inequalities hold from $j=2$ to $j=K_L$ and $j = K_S$ respectively; the $j=1$ is already constrained by \eqref{thirdofmany}, while the $j = 0$ is constrained by \eqref{allequal}, and the $j > K_L$ and $j > K_S$ cases are governed by \eqref{secondofmany} and \eqref{firstofmany} respectively.

As Proposition \ref{NUMprop} shows, the memory inequality constraints allow us to greatly simplify the original objective function \eqref{generalorig}.

\begin{proposition}
\label{NUMprop}
Assuming uniform file length and popularity but two different user
cache sizes, and with variables as defined in \eqref{secondnewvar} and satisfying \eqref{firstofmany}-\eqref{allequal} and \eqref{6ththirdconstraint}-\eqref{6thfifthconstraintp3},  the objective function \eqref{generalorig} simplifies exactly as
 \begin{multline}
 \mathbb{E}\left[\sum_{\mathcal{S}\in \mathcal{P}(\mathcal{U})\setminus \emptyset}\max_{k \in \mathcal{S}}\{|W_{S\setminus\{k\}}^{(d_k)}|\} \right] \\
 = \sum_{j=0}^{K-1} \binom{K_s}{j+1}(v_{j,S}-v_{j,M}) + \binom{K}{j+1}v_{j,M} +
 \\ \quad \left(\binom{K_L}{j+1} +\binom{K_S}{1}\binom{K_L}{j}\right)(v_{j,L}-v_{j,M}) \label{simpobjfunc3}
 \end{multline} 
\end{proposition}

The proof of Proposition \ref{NUMprop} is in Appendix \ref{D}. The simplified optimization problem can then be written as 
\begin{align}
\mbox{minimize} &  \sum_{j=0}^{K-1} \binom{K_s}{j+1}(v_{j,S}-v_{j,M})+ \binom{K}{j+1}v_{j,M} + \nonumber \\
 & \quad \left(\binom{K_L}{j+1} +\binom{K_S}{1}\binom{K_L}{j}\right)(v_{j,L}-v_{j,M}) \label{6thobjfunction} \\
\mbox{subject to} & \text{ \eqref{firstofmany}-\eqref{allequal}, \eqref{6thfirstconstraint}, \eqref{6thsecondconstraintp1}-\eqref{6thfifthconstraintp3}}. \label{6thconstraints}
\end{align}

This is a linear program that has a number of variables, constraints, and objective function summands that scale linearly in $K$. 

Finally, we remark that only two cache sizes were considered in this paper. While a practical system would likely have more than two cache sizes, it is reasonable to expect that only a small number of cache sizes will be used, e.g. cell phones with 8,16, 32 or 64 GB of cache memory. The reasoning used here for two cache sizes could then be extended to accommodate these additional cache sizes as needed.

\subsection{Numerical Results}

\begin{figure}%[htbp!]
\centering
\includegraphics[width=3.6in]{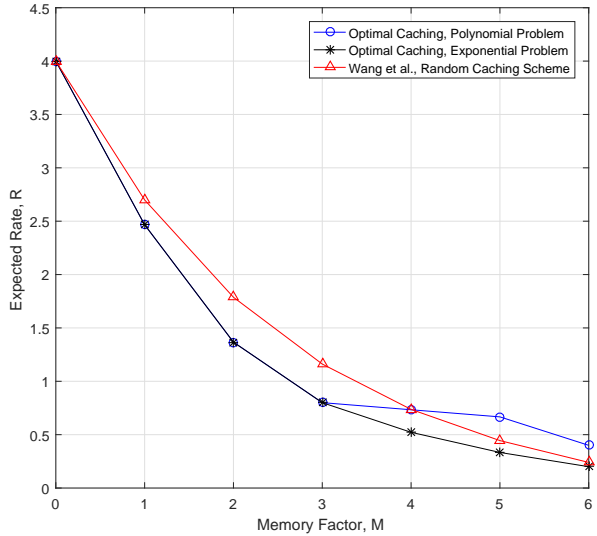}
\caption{\label{NUMpic2} A comparison of the performance of the solution obtained from \eqref{generalorig}-\eqref{generalconstraint} to the solution obtained by \eqref{6thobjfunction}-\eqref{6thconstraints} for $K_S = 2$.}
\end{figure}

Consider a case where there are $N = 6$ files, uniform in
popularity and length, and $K = 4$ users. Define a "memory factor" 
$M \in [0:N]$; %\footnote{Note the slight abuse of notation.} 
there are $K_S$
small users with a cache size of $M_S = 0.8M$, and $K_L$ large users
with a cache size of $M_L = 1.2M$. %\footnote{In some cases, this gives the large users a cache larger than the entire library of files. In this case, the extra cache space provides no extra performance and so it does not matter if the large cache sizes are capped at $M_L$ = N or not.} 
Fig.~\ref{NUMpic2} compares the corresponding solution of the
general problem \eqref{generalorig}-\eqref{generalconstraint} to the
solution of the simplified problem
\eqref{6thobjfunction}-\eqref{6thconstraints} when there are $K_S = 2$
small users. The random caching scheme of \cite{2015arXiv150401123W}
is included, but the centralized scheme of \cite{7925535} (which has
exponential complexity) is not. The
purpose here is not to determine the best caching scheme for the
heterogeneous cache case, but to, first, demonstrate the implicit
performance-tractability tradeoff of using the simplified problem
\eqref{6thobjfunction}-\eqref{6thconstraints} over the general
problem, and second, to demonstrate that it is worth the effort of
developing and using these problems to design cache content (rather
than caching randomly) when the engineering context allows for it.
Nevertheless, we expect the performance of the problem in \cite{7925535} to be very similar, if not identical, to the exponential problem developed here, even though the two optimization frameworks are not identical themselves.
Table \ref{NUMTable} also shows the optimal cache content obtained
from the general problem in the $K_S = 2$, $M = 4$ case. 

%\begin{figure}[htbp!]
%\centering
%\includegraphics[width=3.0in]{Figures2/NonUniformCacheSizePlotFor1Small.eps}
%\caption{\label{NUMpic1} A comparison of the performance of the solution obtained from \eqref{generalorig}-\eqref{generalconstraint} to the solution obtained by \eqref{6thobjfunction}-\eqref{6thconstraints} for $K_S = 1$.}
%\end{figure}

\begin{table}
\centering
\caption{Optimal subfile sizes and memory allocation for the general problem \eqref{generalorig}-\eqref{generalconstraint} with $K=4, N = 6$, $K_S = 2$, $M_S = 3.2$ and $M_L = 4.8$, with values rounded to three decimal places.}
\label{NUMTable}
\begin{tabular}{ | c | c | c | c | c | c | c | } 
\hline
& \multicolumn{6}{|c|}{File Index}\\
\hline
Subset &1&2&3&4&5&6  \\ 
\hline
$\emptyset$ & 0 & 0& 0 & 0& 0& 0 \\ 
\hline
$\dots $& \multicolumn{6}{|c|}{$\dots$}\\
\hline
$\{4\} $& 0 & 0& 0 & 0& 0& 0 \\ 
\hline
$\{1,2 \} $&0.056  &0.056 &0.056 &0.056 & 0.056& 0.056 \\ 
\hline
$\{1,3 \}$ &0.056  &0.056 &0.056 &0.056 & 0.056& 0.056  \\ 
\hline
$\{1,4 \}$&0.056  &0.056 &0.056 &0.056 & 0.056& 0.056\\ 
\hline
$\{2,3 \}$ &0.056  &0.056 &0.056 &0.056 & 0.056& 0.056 \\  
\hline
$\{2,4 \}$&0.056  &0.056 &0.056 &0.056 & 0.056& 0.056\\  
\hline
$\{3,4 \} $&0.056  &0.056 &0.056 &0.056 & 0.056& 0.056\\ 
\hline
$\{1,2,3\}$ & 0.033 &0.033&0.033 & 0.033&0.033& 0.033 \\ 
\hline
$\{1,2,4\} $& 0.033&0.033&0.033 & 0.033&0.033& 0.033 \\ 
\hline
$\{1,3,4\} $& 0.300 &0.300&0.300 & 0.300&0.300& 0.300 \\ 
\hline
$\{2,3,4\}$ & 0.300 &0.300&0.300 & 0.300&0.300& 0.300 \\ 
\hline
$\{1,2,3,4\} $& 0 &0&0 & 0&0& 0 \\
\hline 
Mem. (L): & 0.800 & 0.800& 0.800& 0.800& 0.800 & 0.800 \\
\hline
Mem. (S): & 0.533 & 0.533& 0.533 & 0.533 & 0.533 & 0.533 \\
\hline
\end{tabular}
\end{table} 

%Figure \ref{NUMpic1} shows that in the $K_S = 1$ case, the simplified problem performs identically to the general problem, and although it is not shown here, this is also true in the $K_S = 3$ case. In these cases, the choice of variables given by \eqref{secondnewvar} and the memory inequality constraints \eqref{6thfifthconstraintp1}-\eqref{6thfifthconstraintp3} captured the behaviour of the general problem perfectly.
Fig.~\ref{NUMpic2} shows that in the $K_S = 2$ case, while the
simplified problem tracks the optimal scheme for small cache size
($M\le 3$), it performs worse than even the random caching scheme for
large $M$ values. Table \ref{NUMTable} reveals why this is the case.
Here, users
1 and 2 are the small users, and users 3 and 4 are the large users.
The variable definitions in \eqref{secondnewvar} specify one
variable for all mixed subsets of the same size, but consider the
subfile sizes for the size-three subsets: the subsets with 2 small
users have smaller subfiles than the subsets with 2 large users. Thus
using only one $v_{l,j}^M$ variable for these four subsets results in
a loss in performance. In principle, one could introduce more
variables to accommodate this, albeit at a cost of a more complicated 
objective function. % would have to be derived, but if the number of variables still scales as a polynomial function of $K$ and $N$, then it may be worthwhile for large systems. 

%\footnote{The performance of the optimization problems for $K_S = 1$ and 3 was also considered, although their plots are excluded here for the sake of brevity. In those cases, the simplified problem had identical performance to the general problem, and so the performance degradation of the simplified problem for large $M$ does not occur for all values of $K_S$. }, 

\section{Coded Caching with Heterogeneous Files and Cache Sizes}
\label{NU3}

%\subsection{Background}

The natural final step in this program is to develop a tractable
optimization problem that accommodates non-uniformity in cache size,
file size, and popularity at the same time. To the best of our
knowledge, there has yet to be a caching scheme proposed that handles
heterogeneity in all three of these domains. %; indeed, the only paper we have seen combining any two of them is the previously-mentioned work \cite{7523930}, which characterizes the achievable rate region for non-uniform cache and file size, but only for $K=2$ users and $N=2$ files. 
 
\subsection{Optimization Formulation}
\label{finprobform}
We begin by defining a new set of variables that, in a sense, combines the functionality of the variables defined in \eqref{firstnewvar} and \eqref{secondnewvar}. Let
\begin{equation}
|W^{(l)}_{S}| = 
\begin{cases}
v^S_{l,j} & \text{if $\mathcal{S}$ contains only small-cache users,}  \\
v^L_{l,j} & \text{if $\mathcal{S}$ contains only large-cache users,} \\
v^M_{l,j} & \text{otherwise},  \\
\end{cases}
 \label{thirdnewvar}
\end{equation}
for a subset of users $\mathcal{S}$ such that $|\mathcal{S}| = j$ and all files $l \in [N]$. The symmetry across users is broken by the heterogeneity of the user cache size, and the symmetry across files is broken by a the heterogeneity of files in both length and popularity. Nevertheless, we can still exploit the symmetry across subsets of users of the same size ($j$) and type (i.e. small, large, mixed) for a particular file ($l$) to reduce the complexity of the original problem \eqref{generalorig}-\eqref{generalconstraint} while still accounting for the aforementioned heterogeneity. 

Many of the constraints used in the non-uniform cache size case can be
converted to their equivalents for this new case. If $j>K_S$, there cannot be a subset of $j$ small cache users, so
\begin{equation}
v_{l,j}^S = 0, \forall j > K_S, \forall l \in [N], \label{2ndfirstofmany}
\end{equation}
and similarly
\begin{equation}
v_{l,j}^L = 0, \forall j > K_L, \forall l \in [N].
\end{equation}
Since there cannot be a subset of size 1 containing both large and small users, the $l=1$ variable is constrained as
\begin{equation}
v_{l,1}^M = 0, \forall l \in [N],
\end{equation}
and for convenience, we set
\begin{equation}
v_{l,0}^S = v_{l,0}^L = v_{l,0}^M = v_{l,0}, \forall l \in [N]. \label{zeroeq}
\end{equation}
The cache size-based memory inequalities should still hold, giving, for a fixed file $l$,
\begin{align}
v_{l,j}^{L} \geq &\,\,v_{l,j}^{M},\,\, j \in [2:K_L], \forall l \in [N], \label{7thfifthconstraintp1} \\
v_{l,j}^{M} \geq &\,\,v_{l,j}^{S},\,\, j \in [2:K_S], \forall l \in [N],  \label{7thfifthconstraintp2} \\
v_{l,j}^{L} \geq &\,\,v_{l,j}^{S},\,\, j \in [1:K_L], \forall l \in [N]  \label{7thfifthconstraintp3}. 
\end{align}

We also import conditions from the non-uniform file size and
popularity problem. As seen earlier, it is better to prioritize file
length rather than popularity, and so we label the files in decreasing
order of file length. %see Section \ref{NUPFM4} for further discussion about this choice. 
This gives
\begin{align}
v_{l_1,j}^{L} \geq &\,\,v_{l_2,j}^{L}, \, j \in [0:K-1], \forall l_1,l_2 \text{ s.t. } l_1 < l_2 \label{7thnthconstraintp1} \\
v_{l_1,j}^{M} \geq&\,\, v_{l_2,j}^{M},\, j \in [0:K-1], \forall l_1,l_2 \text{ s.t. } l_1 < l_2  \label{7thnthconstraintp2} \\
v_{l_1,j}^{S} \geq&\,\, v_{l_2,j}^{S}, \, j \in [0:K-1],\forall l_1,l_2 \text{ s.t. } l_1 < l_2  \label{7thnthconstraintp3}. 
\end{align}

The remaining conditions are formed using identical reasoning to the
non-uniform cache case, but occur on a file-by-file basis as needed.
The file reconstruction constraint remains the same, with the minor
change that the file must add up not to the common file length 1, but
to $F_l,$ the actual length of the file as expressed below:
\begin{multline}
F_l = 
\sum_{j=0}^K\binom{K_L}{j}(v_{l,j}^{L}-v_{l,j}^{M}) + \binom{K_S}{j}(v_{l,j}^{S}-v_{l,j}^{M}) \\
+ \binom{K}{j}v_{l,j}^{M},  \qquad \forall l \in [N] \label{7thfirstconstraint}
\end{multline}

The cache memory constraints are modified similarly, except instead of giving an equal amount $M_S/N$ to each file, an amount $\mu_l^S$ is allocated to file $l$, yielding
\begin{multline}
\mu_l^S = \sum_{j=1}^K  \binom{K_S-1}{j-1}(v_{l,j}^{S}-v_{l,j}^{M}) \\
	+\binom{K-1}{j-1}v_{l,j}^{M}, \qquad \forall l \in [N], \label{7thsecondconstraintp1}
\end{multline}
where it must be the case that
\begin{equation}
\sum_{l = 1}^N \mu_{l}^S = M_S .\label{7thmemconstraintp1}
\end{equation}
A similar pair of equations holds for large users:
\begin{multline}
\mu_l^L = \sum_{j=1}^K \binom{K_L-1}{j-1}(v_{l,j}^{L}-v_{l,j}^{M}) \\
+ \binom{K-1}{j-1}v_{l,j}^{M},  \qquad \forall l \in [N], \label{7thsecondconstraintp2}
\end{multline}
and
\begin{equation}
\sum_{l = 1}^N \mu_{l}^L = M_L .\label{7thmemconstraintp2}
\end{equation}
The subfiles are also required to be positive in size, as always:
\begin{eqnarray}
v_{l,j}^L \geq 0,\, v_{l,j}^S \geq 0,\, v_{l,j}^M \geq 0,\, \forall j \in [0:K],\, l \in [N]
\end{eqnarray}

Finally, another set of memory inequality constraints are required to break ties between small-index, small-cache subfiles and large-index, large-cache subfiles. Leaving the justification and discussion of this choice to later sections, we develop a caching scheme under the constraints
\begin{align}
v_{l_1,j}^{L} \geq&\,\, v_{l_2,j}^{M},\,\, j \in [2:K_L],\,\, \forall l_1,l_2 \in [N] \label{7thfinconstraintp1} \\
v_{l_1,j}^{M} \geq&\,\, v_{l_2,j}^{S},\,\, j \in [2:K_S],\,\, \forall l_1,l_2, \in [N] \label{7thfinconstraintp2} \\
v_{l_1,j}^{L} \geq&\,\, v_{l_2,j}^{S},\,\, j \in [1:K_L],\,\, \forall l_1,l_2 \in [N] \label{7thfinconstraintp3}. 
\end{align}

In words, this means that subfiles for any file stored on a larger cache type should be larger than the subfiles of \emph{any} file stored on a smaller cache type, independent of which files are involved. The following proposition gives the objective function of the simplified optimization problem.
\begin{proposition}
\label{finprop}
Define the following functions of the integer parameters $n,m,j$ and $i$:
\begin{eqnarray}
\nu_1(n,m,j,i) &= &\left(\frac{K_S-m}{K-n+1}\right) \nonumber \\
&&\binom{K_S-m-1}{i}\binom{K_L-n+1+m}{j-i}, \nonumber \\
\nu_2(n,m,j,i) &=& \left(\frac{K_L-n+1+m}{K-n+1}\right) \nonumber \\
&&\binom{K_S-m}{i}\binom{K_L-n+m}{j-i},  \nonumber
\end{eqnarray}
and
\begin{eqnarray}
\nu(n,j) & = & \sum_{m = 0}^{n-1} \frac{\binom{K_S}{m}\binom{K_L}{n-1-m}}{\binom{K}{n-1}} \nonumber \\ 
&& \left(\sum_{i=1}^{j-2} \nu_1(n,m,j,i) + \sum_{i=2}^{j-1} \nu_2(n,m,j,i) \right) \nonumber
\end{eqnarray}
Then for the variables defined in \eqref{thirdnewvar} satisfying \eqref{2ndfirstofmany}-\eqref{7thfinconstraintp3}, the objective function \eqref{generalorig} simplifies exactly as
 \begin{align}
 &\mathbb{E}\left[\sum_{\mathcal{S}\in \mathcal{P}(\mathcal{U})\setminus \emptyset}\max_{k \in \mathcal{S}}\{|W_{S\setminus\{k\}}^{(d_k)}|\} \right] \nonumber \\
 &\,\,= \sum_{i=0}^{K-1}\sum_{l = 1}^N \text{Pr}[Y_i = l]v_{l,0}^M \nonumber \\
 &\quad+ \sum_{j=1}^{K_L-1} \sum_{i=0}^{K_L-1} \sum_{l=1}^{N} \binom{K_L-1-i}{j}\text{Pr}[Y_i^L = l]v_{l,j}^L \nonumber \\
 &\quad+ \sum_{j=1}^{K_S-1} \sum_{i=0}^{K_S-1} \sum_{l=1}^{N} \binom{K_S-1-i}{j}\text{Pr}[Y_i^S = l]v_{l,j}^S \nonumber \\
 &\quad + \sum_{j=1}^{K_L} \sum_{i=0}^{K_S-1} \sum_{l=1}^{N} \binom{K_L}{j}\text{Pr}[Y_i^S = l]v_{l,j}^L \nonumber \\
 & \quad + \sum_{j=2}^{K_S-1} \sum_{i=0}^{K_S-1} \sum_{l=1}^{N} \binom{K_S-1-i}{j}\binom{K_L}{1}\text{Pr}[Y_i^S = l]v_{l,j}^M \nonumber \\ 
 &\quad + \sum_{j=\max\{K_S,K_L\}+1}^{K-1} \sum_{i=0}^{K-1} \sum_{l=1}^{N} \binom{K-1-i}{j}\text{Pr}[Y_i = l]v_{l,j}^M \nonumber \\
 & \quad + \sum_{j=3}^{\max\{K_S,K_L\}} \sum_{n=1}^{K} \sum_{l=1}^{N} \text{Pr}[Y_{n-1}=l]\nu(n,j)v_{l,j}^M \label{finobjfunc} 
\end{align} 
\end{proposition}
Proposition \ref{finprop} is proved in Appendix \ref{E}. Although visually complicated, \eqref{finobjfunc} simplifies the original objective function \eqref{generalorig} by using only $3(K+1)N$ variables and having a number of terms that scales with $K^2N$ rather than $(2N)^K$. The number of constraints described by \eqref{2ndfirstofmany}-\eqref{7thfinconstraintp3} scales with $KN^2$, and so the following optimization problem is a tractable method of obtaining a caching scheme that accommodates heterogeneity in cache size, file size, and file popularity:
\begin{align}
\mbox{minimize} \;\;\; & \eqref{finobjfunc}  \label{finorig} \\
\mbox{subject to} \;\;\; & \eqref{2ndfirstofmany}-\eqref{7thfinconstraintp3} \label{finconstraints}
\end{align}

\subsection{Numerical Results}

\begin{figure}%[htbp!]
\centering
\includegraphics[width=3.6in]{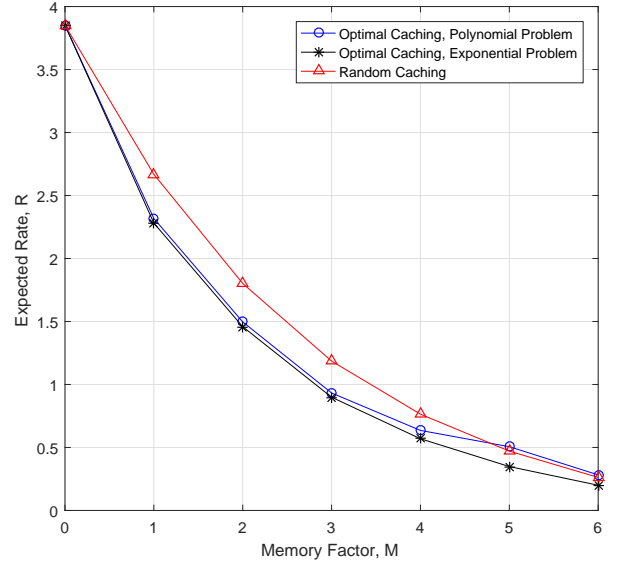}
\caption{\label{NUMPFpic} A comparison of the performance of the solution obtained from \eqref{generalorig}-\eqref{generalconstraint} to the solution obtained by \eqref{finorig}-\eqref{finconstraints}, with reference to a baseline random caching scheme, for the case of non-uniform file size, file popularity, and cache size.}
\end{figure}

To demonstrate that this simplified optimization problem performs well
when compared to the general problem
\eqref{generalorig}-\eqref{generalconstraint}, we again consider $K=4$
users ($K_S = 2$ of which are small-cache users) and $N=6$ files, with
all of the non-uniformities used thus far: file labels and
popularity/size pairs as given in Table \ref{NUPFOrder1}, and cache
sizes of $M_S = 0.8M$ and $M_L = 1.2M$ for $M \in [0:N]$.
Fig.~\ref{NUMPFpic} compares the simplified and general problems with
a naive random caching baseline.

\begin{figure}%[htbp!]
\centering
\includegraphics[width=3.6in]{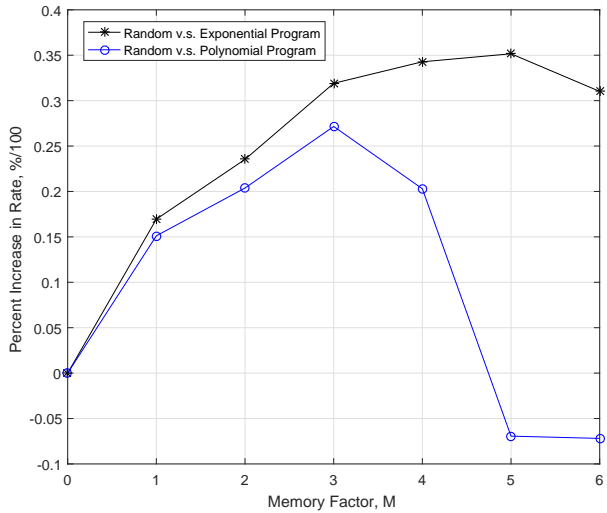}
\caption{\label{NUMPFpic2} The percent increase in expected rate due to random cache content when compared to the simplified problem \eqref{finorig}-\eqref{finconstraints} and general problem \eqref{generalorig}-\eqref{generalconstraint} optimal solutions. }
\end{figure}

We see that the simplified problem yields a scheme that closes
mirrors, although does not match exactly, the performance of the
scheme obtained from the general problem for small and intermediate
$M$ values. Table \ref{NUMPFTable} shows the optimal solution to the
general problem for the $M=2$ case; we see several violations of the
memory inequality constraints that explain why the simplified problem
could not achieve as good a performance as the general problem: the
``true'' optimal solution lies outside of its feasible space.
Nevertheless, the simplified problem still achieves good performance
compared to the general problem in this regime. For large $M$, we see
that the expected rate of the simplified scheme does not drop as
quickly as the general problem solution, and is even eclipsed by the
random caching scheme. This occurs for the same reason we saw in
Section \ref{NUM} in the $K_S = 2$ case (Fig.~\ref{NUMpic2}). To
compare the relative performance of the three schemes in general,
Fig.~\ref{NUMPFpic2} shows the percent increase in expected rate if the random caching scheme is used over the general and simplified schemes respectively. The significant increase in rate when using random caching make it clear that designing the cache content can be worthwhile when the engineering context allows for it.

\begin{table}
\centering
\caption{Optimal subfile sizes and memory allocation for the general problem \eqref{generalorig}-\eqref{generalconstraint} with $K=4, N = 6$, $K_S = 2$, $M_S = 1.6$, $M_L = 2.4$ and file popularity/ size pairs given by Table \ref{NUPFOrder1}, with values rounded to three decimal places.}
\label{NUMPFTable}
\begin{tabular}{|c|c|c|c|c|c|c|}
\hline
& \multicolumn{6}{|c|}{File Index}\\
\hline
Subset &1&2&3&4&5&6  \\ 
\hline
$\emptyset$ & 0 & 0& 0 & 0& 0& 0 \\ 
\hline
$\{1\} $& 0.178 & 0.178& 0.178 & 0.178& 0.165& 0.085 \\ 
\hline
$\{2\}$ & 0.178 & 0.178& 0.178 & 0.178& 0.165& 0.085 \\ 
\hline
$\{3\}$ & 0.178 & 0.178& 0.178 & 0.178& 0.165& 0.085 \\ 
\hline
$\{4\} $& 0.178 & 0.178& 0.178 & 0.178& 0.165& 0.085 \\ 
\hline
$\{1,2 \} $&0&0.077 &0.008 &0& 0& 0 \\ 
\hline
$\{1,3 \}$ &0.090  &0.090  &0.090  &0.008 & 0& 0  \\ 
\hline
$\{1,4 \}$&0.090  &0.090  &0.090  &0.008 & 0& 0\\ 
\hline
$\{2,3 \}$ &0.090  &0.090  &0.090  &0.008 & 0& 0 \\  
\hline
$\{2,4 \}$&0.090  &0.090  &0.090  &0.008 & 0& 0\\  
\hline
$\{3,4 \} $&0.431 &0.188 &0.090 &0.090 & 0.008& 0\\ 
\hline
$\{1,2,3\}$ & 0 & 0& 0 & 0& 0& 0  \\ 
\hline
$\dots$& \multicolumn{6}{|c|}{$\dots$}\\
\hline
$\{1,2,3,4\} $& 0 &0&0 & 0&0& 0 \\
\hline 
Mem. (L): & 0.788 & 0.554& 0.446& 0.284& 0.173 & 0.165 \\
\hline
Mem. (S): & 0.357 & 0.434& 0.365 & 0.194 & 0.165 & 0.085 \\
\hline
\end{tabular}
\end{table}

\subsection{Further Extensions}
\label{NUPFM4}
We first echo the earlier comments about heterogeneous cache sizes: we
consider only two different cache sizes here, but it is possible to
use the same reasoning to develop a tractable optimization problem for
a practical system having more (but not many more) cache sizes. 

The primary focus in this section, however, is on the memory
inequality constraints
\eqref{7thfinconstraintp1}-\eqref{7thfinconstraintp3} used in
developing the simplified problem
\eqref{finorig}-\eqref{finconstraints} of this section. Recall that
these constraints require, among other things, that (roughly speaking)
for a fixed index $j$, the $v^L_{l,j}$ variables for \emph{all} files
$l$ be larger than the $v^S_{l,j}$ variables for all files. Thus the
large-user subfile for the smallest file is larger than the small-user
subfile for the largest file. This restriction was required to allow
us to write the simplified problem, and numerical results show that the
simplified problem still performed well compared to the general
problem for the considered parameters. While a full investigation
of the performance of the simplified problem across all parameter
values would be labourious, it is still possible to
estimate the behaviour for certain parameter regimes. We should expect
\eqref{7thfinconstraintp1}-\eqref{7thfinconstraintp3} to result in a
good simplification when the disparity between the large and small
cache sizes is big, and when there are small numbers of files, because
the large cache users will likely store much larger subfiles than the
small cache users, irrespective of the length of the file. Conversely,
we should expect the performance of
\eqref{finorig}-\eqref{finconstraints} to be relatively poor when the
cache sizes are comparable and there are large numbers of files. In
this case, it may make more sense to use something like the
``opposite'' memory inequality constraint: a small-user subfile
variable $v^S_{l_1,j}$ should be larger than any larger-user subfile
variable $v^L_{l_2,j}$ if file $l_1$ is larger than file $l_2$. While
the simplified optimization problem that would result from this
constraint is not explored in this paper, it should be possible to
construct such a problem using the same kind of reasoning used here. 

Indeed, there may also be other memory inequality constraints that prove to yield useful simplified optimization problems for other parameter sets. The appeal of the tractability of these models is that a server, knowing the relevant parameters for its system, could easily compute the performance of these schemes and choose the best among them; any discussion of the specifics of such schemes, however, is left to future work.

\section{Summary and Conclusions}
\label{Conclusion}
The two primary goals of this paper are to advance a certain
optimization theoretic approach to coded caching problems, and to use
that framework to derive both specific caching schemes and general
insight for system models containing multiple heterogeneities that
have yet to be considered in the literature. An exponentially-scaling
optimization problem corresponding to a caching scheme capable of
handling non-uniform file size, popularity, and cache size is developed. 
It is shown that the original scheme of Maddah-Ali and Niesen in
\cite{6620392, 6763007} is the optimal solution of that problem for
the special case of uniform file length, popularity, and cache size.

Tractable problems are then developed to handle various combinations
of heterogeneous system parameters. The consideration of these special
cases also permitted the observation of the effects that these
non-uniformities have on the optimal cache content. 
When considering non-uniform file popularity and size jointly,
it is shown that
while popularity may in general have some influence on the optimal cache
allocation, file size can be a much stronger influence; indeed, very
good performance is obtained in the case considered by ignoring file
popularity altogether. Finally, with the insights obtained from the
previously-explored special cases, we developed a tractable
optimization problem corresponding to a caching scheme capable of
accommodating all three of the aforementioned heterogeneities, and
showed numerically that it performs well compared to the original
exponentially-scaling problem.

%Future work should explore and characterize the limits of the performance of the tractable caching schemes presented here. While these schemes were shown to perform well under the specific conditions in this paper, the set of possible parameter combinations is very large. While, as discussed earlier, the tractability of the problems allows for the possibility that a server could compute the results of several different problems and use the scheme that performs best, it would be preferable to have analytic (or at least strong numerical) results that would facilitate comparison between schemes without solving the optimization problem. Nevertheless, the various insights and caching schemes obtained from the optimization framework advanced here, taken jointly with the results obtained other work using similar optimization methods, i.e. \cite{7925535} and \cite{2017arXiv170707146J}, provide strong evidence that such optimization frameworks will be of great utility in the design of future caching schemes and in the analysis of future caching problems. Other future directions could include the development of information theoretic outer bounds for the various heterogeneous cases considered here, and the inclusion of even more heterogeneous effects, like non-uniform user download rates.

%--------------------------------------------------------------------------------------------------------------------------------------
%--------------------------------------------------------------------------------------------------------------------------------------

\appendix
 
\subsection{Proof of Lemma \ref{problemma}}
\label{B}

We begin with the proof of the expression of $\text{Pr}[Y_0 = i]$, for
which we use induction on $i$ for $i = 1, \dots, N$. We begin first
with the $i = 1$ case. Let $\mathbf{Z} \in [N]^K$ denote the sequence of outcomes from the $N$ trials, e.g. if for $K$ = 3 and $N$ = 4, trial 1 obtains outcome 2, trial 2 obtains outcome 4, and trial 3 obtains outcome 1, we have $\mathbf{Z} = [2, 4, 1]^T$. 
 Let $X_n$ denote the random variable
representing the number of times the outcome $n$ occurs in the $K$ trials (i.e. the number of times it appears in $\mathbf{Z}$), and stack the $X_n$ variables in a vector $\mathbf{X} \in [K]^N$; the example above would yield $\mathbf{X} = [1, 1, 0, 1]^T$.
% in the multinomial random vector $\mathbf{X}$. 
Then the smallest element of $\mathbf{Z}$ is 1 (i.e. $Y_0 = 1$) if and only if $X_1 >= 1$; in other words, since 1 is the smallest possible outcome, if it occurs anywhere in $\mathbf{Z}$ then it is the smallest element. Thus we have 
\begin{eqnarray}
\text{Pr}[Y_0 = 1] &=& \text{Pr}[X_1 \geq 1] = 1 - \text{Pr}[X_1 = 0]  \nonumber \\
 &=& 1 - (1-p_1)^K   =  \left(\sum_{l=1}^{N}p_l\right)^K - \left(\sum_{l=2}^{N} p_l\right)^K, \nonumber
\end{eqnarray}
which is indeed the formula \eqref{m0case} with $i=1$, as desired.

For an arbitrary $i$ such that $2 \leq i \leq N-1$, we note that if the smallest element of $\mathbf{Z}$ is $i$, (i.e. $Y_0=i$), then there cannot be any values smaller than $i$, and there must be at least one $i$ in $\mathbf{Z}$; in other words:
\begin{align}
\text{Pr}[Y_0 = i] &= \text{Pr}[X_i \geq 1, X_{i-1} = 0, \dots X_1 = 0] \nonumber \\
	         & =  \text{Pr}[X_1 = 0]\text{Pr}[X_2 = 0 | X_1 = 0] \cdots \nonumber \\
	         & \quad \text{Pr}[X_{i-1} = 0 | X_{i-2} = \dots =X_{1} = 0] \nonumber \\
	         & \quad \text{Pr}[X_{i} \geq 1 | X_{i-1} =  \dots = X_{1} = 0] \nonumber \\
	         & =  \text{Pr}[X_1 = 0]\text{Pr}[X_2 = 0 | X_1 = 0] \cdots \nonumber \\
	         & \quad \text{Pr}[X_{i-1} = 0 | X_{i-2} = \dots =X_{1} = 0] \nonumber \\
	         & \quad \left(1-\text{Pr}[X_{i} = 0 | X_{i-1} =  \dots = X_{1} = 0]\right). 
	         \label{verylongexpr} 
\end{align}
Now, comparing the expression \eqref{verylongexpr} for $Y_0 = i$ to the same expression with $Y_0 = i-1$, it is easy to show that
\begin{align}
\text{Pr}[Y_0 = i] &= \text{Pr}[Y_0 = i-1]  \nonumber \\
& \quad \left(\frac{\text{Pr}[X_{i-1} = 0| X_{i-2} = \dots = X_{1} = 0]}{1-\text{Pr}[X_{i-1} = 0| X_{i-2} = \dots =X_{1} = 0]}\right) \nonumber \\
& \quad \left(1 - \text{Pr}[X_i = 0 | X_{i-1} = \dots = X_{1} = 0] \right). \label{alsolongexpr}
\end{align} 

It is possible to compute these conditional probabilities directly:
\begin{eqnarray}
\text{Pr}[X_j = 0 | X_{j-1} = \dots = X_1 = 0] = \left(1 - p_j^{(j)}\right)^K, \label{condprobcomp}
\end{eqnarray}
where 
\begin{equation}
p_j^{(j)} = \frac{p_j}{\sum_{l=j}^{N}p_l} \nonumber
\end{equation}
is the probability of outcome $j$ occurring in a trial conditioned on the knowledge that outcomes 1 through $j-1$ have not occurred in that trial. From the definition of $p_j^{(j)}$, we can rewrite \eqref{condprobcomp} as
\begin{equation}
\text{Pr}[X_j = 0 | X_{j-1} = \dots = X_1 = 0] = \left( \frac{\sum_{l=j+1}^{N}p_l}{\sum_{l=j}^{N}p_l} \right)^K. \nonumber
\end{equation}
Evaluating this expression for $j = i$ and $j = i-1$, we can obtain from \eqref{alsolongexpr}, after some mild algebraic manipulation,
\begin{align}
\text{Pr}[Y_0 = i] =& \,\,\text{Pr}[Y_0 = i-1] \left( \frac{\sum_{l=i}^{N}p_l}{\sum_{l=i-1}^{N}p_l} \right)^K \nonumber \\
& \frac{ \left(\sum_{l=i-1}^{N}p_l\right)^K }{ \left(\sum_{l=i-1}^{N}p_l\right)^K - \left(\sum_{l=i}^{N}p_l\right)^K } \nonumber \\
& \frac{\left(\sum_{l=i}^{N}p_l\right)^K - \left(\sum_{l=i+1}^{N}p_l\right)^K}{\left(\sum_{l=i}^{N}p_l\right)^K} \nonumber \\
 = & \,\,\text{Pr}[Y_0 = i-1] \frac{\left(\sum_{l=i}^{N}p_l\right)^K - \left(\sum_{l=i+1}^{N}p_l\right)^K}{\left(\sum_{l=i-1}^{N}p_l\right)^K - \left(\sum_{l=i}^{N}p_l\right)^K }. \nonumber \\ \label{induchyp2}
\end{align}
Now by the inductive hypothesis, $\text{Pr}[Y_0 = i-1 ]$ is precisely equal to the denominator of \eqref{induchyp2}, and so we obtain
\begin{equation}
\text{Pr}[Y_0 = i] = \left(\sum_{l=i}^{N}p_l\right)^K - \left(\sum_{l=i+1}^{N}p_l\right)^K,
\end{equation} 
as desired.

Although the $\text{Pr}[Y_0 = N]$ formula was covered in the preceding paragraph, we discuss it in further detail here because \eqref{m0case} above may not appear sensible in the $i = N$ case. Note that if $N$ is the smallest value in the multinomial vector $\mathbf{Z}$, then it must be the case that every element of $\mathbf{Z}$ is equal to $N$, otherwise some element not equal to $N$ would be the smallest value. Thus we have
\begin{eqnarray}
\text{Pr}[Y_0 = N] & = & \text{Pr}[X_N = K, X_{N-1} = 0, \dots, X_{1} = 0] \nonumber \\
 & = & p_N^K \nonumber \\
 %& = & \left(\sum_{l=N}^{N}p_l\right)^K - 0^K \nonumber \\
 & = & \left(\sum_{l=N}^{N}p_l\right)^K - \left(\sum_{l = N+1}^{N} p_l\right)^K, \nonumber 
\end{eqnarray} 
 which is the formula \eqref{m0case} with $i=N$, noting that we use the definition that $\sum_{l = a}^{b} n_l = 0$ when $a > b$.
 
 Next, we proceed to the $m = 1$ case. Here we will directly compute $\text{Pr}[Y_1  = i]$ by first deriving $\text{Pr}[Y_1 = i , Y_0 = j]$, and then obtaining the desired quantity from the sum
 \begin{equation}
 \text{Pr}[Y_1 = i] = \sum_{j = 1}^{N} \text{Pr}[Y_1 = i, Y_0 = j] \label{usejointprobs}
 \end{equation}
Clearly the second smallest element of $\mathbf{Z}$ is no smaller than the smallest element of $\mathbf{Z}$, so there are two cases to consider: $i > j$, and $i = j$. 
%If $i < j$, then and so the probability of this occurring is 0. 
If $i > j$, we write
\begin{align}
&\text{Pr}[Y_1 = i, Y_0 = j] \nonumber \\
&\,\,=\text{Pr}[X_i \geq 1, X_{i-1} = 0, \dots, X_{j} = 1, \nonumber \\
&\quad \qquad X_{j-1} = \dots = X_{1} = 0] \nonumber \\
&\,\,=\text{Pr}[X_{i-1} = 0, \dots, X_{j} = 1, X_{j-1} = \dots = X_{1} = 0] \nonumber \\
&\quad - \text{Pr}[X_i = 0, X_{i-1} = 0, \dots, X_{j} = 1, \nonumber \\
&\quad \qquad X_{j-1} = \dots = X_{1} = 0] 
\label{bigdiff} 
\end{align} 
 To compute the difference \eqref{bigdiff}, note that we can form a ($K$, 4) sequential vector of outcomes $\tilde{\mathbf{Z}}(m,n)$, indexed by two integers $m$ and $n$, from the original ($K$, $N$) sequential vector of outcomes $\mathbf{Z}$ in the following way: for a single trial, the first outcome of $\tilde{\mathbf{Z}}(m,n)$ occurs if any of the first $n-1$ outcomes of $\mathbf{Z}$ occur, and so it has the probability $\tilde{p}_1 = \sum_{l=1}^{n-1} p_j$; the second outcome of $\tilde{\mathbf{Z}}(m,n)$ occurs if the $n$-th outcome of $\mathbf{Z}$ occurs, and so it has a probability of $\tilde{p}_2 = p_n$; the third outcome of $\tilde{\mathbf{Z}}(m,n)$ occurs if any outcomes of $\mathbf{Z}$ from $n+1$ to $m$ occurs, and so it has a probability of $\tilde{p}_3 = \sum_{l = n+1}^{m}p_l$; and the fourth outcome of $\tilde{\mathbf{Z}}(m,n)$ occurs if any of the last $N-m$ outcomes of $\mathbf{Z}$ occur, and so it has a probability $\tilde{p}_4 = \sum_{l=m+1}^{N}p_l$. If we define $\tilde{X}_j$ as the number of times outcome $j$ occurred in $\tilde{\mathbf{Z}}(m,n)$, then we consequently have $\tilde{X}_1 = \sum_{l=1}^{n-1} X_l$, $\tilde{X}_2 = X_n$, $\tilde{X}_3 = \sum_{l=n+1}^{m} X_l$,  and $\tilde{X}_4 = \sum_{l=m+1}^{N} X_l$. We can then rewrite\footnote{Note that the $\tilde{X}$ variables lose the $(m,n)$ indices of the original variable $\tilde{\mathbf{X}}(m,n)$. This is done for notational convenience, but will result in an abuse of the notation when multiple $\tilde{\mathbf{X}}(m,n)$ are involved. We will therefore be careful to indicate which $\tilde{X}$ variables belong to which $\tilde{\mathbf{X}}(m,n)$ vectors.} \eqref{bigdiff} using vectors $\tilde{\mathbf{X}}(i,j)$ and $\tilde{\mathbf{X}}(i-1,j)$ as
\begin{multline}
\text{Pr}[Y_1 = i, Y_0 = j] \\
= \text{Pr}[\tilde{X}_4 = K-1, \tilde{X}_{3} = 0, \tilde{X}_{2} = 1, \tilde{X}_1 = 0] \\
- \text{Pr}[\tilde{X}_4 = K-1, \tilde{X}_{3} = 0, \tilde{X}_{2} = 1, \tilde{X}_1 = 0]  \nonumber 
\end{multline}  
 where the first term is computed with respect to $\tilde{\mathbf{X}}(i-1,j)$, and the second with respect to $\tilde{\mathbf{X}}(i,j)$. Using the probabilities defined earlier, this gives
\begin{align}
 &\text{Pr}[Y_1 = i, Y_0 = j] \nonumber \\
 &\,\, = K\tilde{p}_2(\tilde{p}_4)^{K-1} - K\tilde{p}_2(\tilde{p}_4)^{K-1} \label{confusing1} \\
 &\,\, = Kp_{j}\left(\left(\sum_{l=i}^{N}p_l\right)^{K-1} - \left(\sum_{l=i+1}^{N}p_l\right)^{K-1}\right) \label{jointprob1}
 \end{align}
 where, again, the terms in \eqref{confusing1} are computed with respect to $\tilde{\mathbf{X}}(i-1,j)$, and $\tilde{\mathbf{X}}(i,j)$ respectively. 

Similar reasoning yields the value of the joint probability when $i = j$:
\begin{align}
&\text{Pr}[Y_1 = i, Y_2 = i] \nonumber \\
&\,\, = \text{Pr}[X_i \geq 2, X_{i-1} = \dots = X_1 = 0] \nonumber \\
&\,\, = \text{Pr}[X_{i-1} = \dots = X_1 = 0] - \text{Pr}[X_i = 0 = \dots = X_1 = 0] \nonumber \\
&\quad -\text{Pr}[X_i =1, X_{i-1} = \dots = X_1 = 0] \nonumber\\
&\,\, = \text{Pr}[\tilde{X}_4 = K, \tilde{X}_3 = \tilde{X}_2 = \tilde{X}_1 = 0] \nonumber \\
& \quad -\text{Pr}[\tilde{X}_4 = K, \tilde{X}_3 = \tilde{X}_2 = \tilde{X}_1 = 0] \nonumber \\
& \quad -\text{Pr}[\tilde{X}_4 = K-1, \tilde{X}_3 = 1,\tilde{X}_2 = \tilde{X}_1 = 0]. \label{conf2}
\end{align}
 Here, the first term in \eqref{conf2} is computed with respect to $\tilde{\mathbf{X}}(i-1,i-2)$, the second with respect to $\tilde{\mathbf{X}}(i,i-1)$, and the third with respect to $\tilde{\mathbf{X}}(i,i-1)$, although this choice of $\tilde{\mathbf{X}}$ variables is not unique. This gives
 \begin{align}
& \text{Pr}[Y_1 = i, Y_2 = i] \nonumber \\
&\,\, = (\tilde{p}_4)^K - (\tilde{p}_4)^K - K\tilde{p}_2(\tilde{p}_4)^{K-1} \nonumber \\
&\,\, = \left(\sum_{l = i}^{N}p_l\right)^K - \left(\sum_{l=i+1}^{N}p_l\right)^K -Kp_i\left(\sum_{l=i+1}^{N}p_l\right)^{K-1}.  \label{jointprob2}
 \end{align}
 We can now evaluate \eqref{usejointprobs} as
 \begin{align}
& \text{Pr}[Y_1 = i] \nonumber \\
 &\,\, = \sum_{j = 1}^{N} \text{Pr}[Y_1 = i, Y_0 = j] \nonumber \\
 &\,\, = \sum_{j = 1}^{i-1} \text{Pr}[Y_1 = i, Y_0 = j] + \text{Pr}[Y_1 = i, Y_0 = i] + 0 \nonumber \\
 &\,\, =\sum_{j = 1}^{i-1}\left( Kp_j\left( \left(\sum_{l=i}^{N}p_l\right)^{K-1} - \left(\sum_{l=i+1}^{N}p_l\right)^{K-1}  \right) \right) \nonumber \\
 & \quad + \left(\sum_{l = i}^{N}p_l\right)^K - \left(\sum_{l=i+1}^{N}p_l\right)^K -Kp_i\left(\sum_{l=i+1}^{N}p_l\right)^{K-1} \nonumber \\
 &\,\, =\left(\sum_{l = i}^{N}p_l\right)^K - \left(\sum_{l=i+1}^{N}p_l\right)^K -Kp_i\left(\sum_{l=i+1}^{N}p_l\right)^{K-1} \nonumber \\
 & \quad + K\left(\sum_{l=i}^{N}p_l\right)^{K-1}\left(\sum_{j=1}^{i-1}p_j\right) \nonumber \\
 & \quad - K\left(\sum_{l=i+1}^{N}p_l\right)^{K-1}\left(\sum_{j=1}^{i-1}p_j\right)  \nonumber \\ 
 & \,\, =\text{Pr}[Y_0 = i] +  K\left(\sum_{l=i}^{N}p_l\right)^{K-1}\left(\sum_{j=1}^{i-1}p_j\right) \nonumber \\
 & \quad - K\left(\sum_{l=i+1}^{N}p_l\right)^{K-1}\left(\sum_{j=1}^{i}p_j\right),
 \end{align}
 which is the desired formula. Note that, although we refer to $\text{Pr}[Y_0 = i]$ in the formula for $\text{Pr}[Y_1 = i]$, this is only for notational simplicity; we do not wish to suggest some sort of interpretation relating the two quantities in this way.
 
 Finally, we must compute $\text{Pr}[Y_m = i]$ for $m = 2, \dots, K-1$. We take an approach similar to the $\text{Pr}[Y_1 = i]$ case, and derive $\text{Pr}[Y_m =i]$ using the joint probabilities $\text{Pr}[Y_m = i, Y_{m-1} = j]$. As before, there are two cases, $i > j$, and $i = j$, as the $m$-th smallest element of $\mathbf{Z}$ cannot be smaller than the $(m-1)$-th element of $\mathbf{Z}$, and so $\text{Pr}[Y_m = i, Y_{m-1} = j] = 0$ if $i < j$. 

In the case where $i > j$, we have
 \begin{align}
 &\text{Pr}[Y_m = i, Y_{m-1} = j] \nonumber \\ 
 &\,\,=\sum_{k=1}^{m} \text{Pr}[X_i \geq 1, X_{i-1} = \dots = X_{j+1} = 0, \nonumber \\
 &\quad  X_j = k,  X_{j-1} + \dots + X_{1} = m-k ] \nonumber \\
 &\,\,=\sum_{k=1}^{m} \text{Pr}[X_{i-1} = \dots = X_{j+1} = 0, X_j = k, \nonumber \\
 &\quad X_{j-1} + \dots + X_{1} = 0]  - \text{Pr}[X_{i} = \dots = X_{j+1} = 0, \nonumber \\
 &\quad X_{j} = k, X_{j-1} + \dots + X_1 = m-k ] \nonumber \\ 
 &\,\,=\left(\text{Pr}[X_{i-1} = \dots = X_{j+1} = 0, X_j + \dots + X_1 = m] \right. \nonumber \\
 &\quad - \left. \text{Pr}[X_{i-1} = \dots = X_{j} = 0, X_{j-1} + \dots + X_{1} = m ] \vphantom{\text{Pr[]}}\right)  \nonumber \\ 
 &\quad - \left(\text{Pr}[X_{i} = \dots = X_{j+1} = 0, X_j + \dots + X_1 = m] \right. \nonumber \\
 &\quad - \left. \text{Pr}[X_i = \dots = X_{j} = 0, X_{j-1} + \dots + X_{1} = m ] \right).  \label{biglongone}
 \end{align} 
Now we recast the four terms of \eqref{biglongone} in terms of $\tilde{\mathbf{X}}(i-1,j+1)$, $\tilde{\mathbf{X}}(i-1,j)$, $\tilde{\mathbf{X}}(i,j+1)$, and $\tilde{\mathbf{X}}(i,j)$ respectively:
\begin{align}
 &\text{Pr}[Y_m = i, Y_{m-1} = j] \nonumber \\ 
 &\,\, =\Big(\text{Pr}[\tilde{X}_4 = K-m,  \tilde{X}_{3} = 0, \tilde{X}_2 = 0,  \tilde{X}_1 = m] \nonumber \\
 &\quad - \text{Pr}[\tilde{X}_{4} = K-m,  \tilde{X}_{3} = 0, \tilde{X}_{2} = 0, \tilde{X}_1 = m ] \Big) \nonumber \\
 &\quad - \Big(\text{Pr}[\tilde{X}_{4} = K-m, \tilde{X}_{3} = 0, \tilde{X}_2 = 0, \tilde{X}_1 = m] \nonumber \\
 &\quad -  \text{Pr}[ \tilde{X}_{4} = K-m, \tilde{X}_{3} = 0, \tilde{X}_2 = 0, \tilde{X}_1 = m ] \Big). \nonumber \\
&\,\, =\binom{K}{K-m}(\tilde{p}_4)^{K-j}(\tilde{p}_1)^j + \binom{K}{K-m}(\tilde{p}_4)^{K-j}(\tilde{p}_1)^j \nonumber \\
&\quad + \binom{K}{K-m}(\tilde{p}_4)^{K-j}(\tilde{p}_1)^j + \binom{K}{K-m}(\tilde{p}_4)^{K-j}(\tilde{p}_1)^j \nonumber \\
&\,\,=\binom{K}{K-m}\left(  \left(\sum_{l=i}^N p_l\right)^{K-m}\left(\sum_{l=1}^{j}p_l\right)^m \right. \nonumber \\
&\quad - \left(\sum_{l=i}^{N}p_l\right)^{K-m}\left(\sum_{l=1}^{j-1}p_l\right)^{m} \nonumber \\
&\quad - \left(\sum_{l=i+1}^{N}p_l\right)^{K-m}\left(\sum_{l=1}^{j}p_l\right)^{m} \nonumber \\ 
&\quad + \left. \left(\sum_{l=i+1}^{N}p_l\right)^{K-m}\left(\sum_{l=1}^{j-1}p_l\right)^{m} \right)  \nonumber \\
&\,\,=\binom{K}{K-m}\left( \left(\sum_{l=i}^N p_l\right)^{K-m} - \left(\sum_{l=i+1}^N p_l\right)^{K-j}  \right) \nonumber \\
& \quad  \left( \left(\sum_{l=1}^{j}p_l\right)^{m} - \left(\sum_{l=1}^{j-1}p_l\right)^{m}  \right).
\end{align} 

In the case where $i = j$, there are two sub-cases to consider: $i = j \neq 1$ and $i = j = 1$. In the former sub-case, we must have $X_i \geq 2$, and $X_{i-1} + \dots + X_{1} = b \leq m-1$. Suppose that $X_{i} = 2 + k$ for some integer $k \in \{0, \dots, K-2\}$. We know that $Y_m = i$ and $Y_{m-1} = i$, but that leaves $k$ $Y$ variables ``adjacent" to $Y_m$ and $Y_{m-1}$ that must also have a value of $i$. Let $n_l$ denote the number of variables $Y_{m'}$ that are equal to $i$ and have $m' > m$, and $n_s$ denote the number of variables $Y_{m'}$ that are equal to $i$ and have $m' < m-1$. Then $n_l + n_s = k$ and the following must be true: there are at most $K-1-m$ variables $Y_{m'}$ with $m' > m$, because there are only $K-1$ total $Y_{m'}$ variables, and so $n_l \leq K-1-m$; moreover there are only $m-2$ variables $Y_{m'}$ with $m' < m-1$, and so $n_s \leq m-2$. We will use these inequalities to place bounds on $b$ as a function of $k$.

We first consider an upper bound on $b$. We have already seen that $b \leq m-1$ in general, but the inequality on $n_l$ induces a second upper bound on $b$ that is sometimes stricter than the first. Note that $b = m-1$ only if $Y_{m-1}$ is the first $Y_{m'}$ variable with the value $i$; if $X_{i} = 2+k$, then we must have $n_l = k$, and therefore $k \leq K-1-m$. Thus if $k  > K-1-m$, then $b < m-1$, where the maximum possible $b$ decreases by one every time $k$ increases by one. Indeed, the upper limit on $b$ is imposed by $(m-1)- (k - (K-1-m)) = K-k-2$; the general upper limit on be is then $b \leq \min\{m-1, K-k-2\}$. 

The lower limit on $b$ is obtained through similar reasoning, but we first note that the trivial lower limit on $b$ is 0, which occurs when the number $i$ constitutes (at least) the first $m$ smallest values of $\mathbf{Z}$; in this case $k \geq m-1$. If $k < m-1$, then not all $Y_{m'}$ with $m'<m$ can have values of $i$. In general, $k+2+b \geq m+1$, which implies that $b \geq m-1-k$. Then general lower bound on $b$ is therefore $b \geq \max\{0, m-1-k\}$.

We are therefore now in a position to write
\begin{align}
&\text{Pr}[Y_m = i, Y_{m-1} = i \neq 1] \nonumber \\
&\,\,= \sum_{k = 0}^{K-2}\sum_{b = \max\{0,m-1-k\}}^{\min\{m-1,K-2-k\}} \nonumber \\
&\quad \text{Pr}[X_i = 2+k, X_{i-1} + \dots + X_1 = b] \nonumber \\
&\,\,= \sum_{k = 0}^{K-2}\sum_{b = \max\{0,m-1-k\}}^{\min\{m-1,K-2-k\}} \nonumber \\
&\quad \text{Pr}[\tilde{X}_4 = K-k-2-b, \tilde{X}_3 = 0, \tilde{X}_2 = 2+k, \tilde{X}_{1} = b]  
\label{beenherebefore}
\end{align}
where the $\tilde{X}$ variables in \eqref{beenherebefore} are with reference to $\tilde{\mathbf{X}}(i,i)$. This can be computed as
\begin{align}
&\text{Pr}[\tilde{X}_4 = K-k-2-b, \tilde{X}_3 = 0, \tilde{X}_2 = 2+k, \tilde{X}_{1} = b] \nonumber \\
&\,\,= \binom{K}{K-k-2-b,b,2+k}(\tilde{p}_4)^{K-k-2}(\tilde{p}_2)^{2+k}(\tilde{p}_1)^b \nonumber \\
&\,\,= \binom{K}{K-k-2-b,b,2+k} \nonumber \\
& \quad \left(\sum_{l=i+1}^{N}p_l\right)^{K-k-2-b}(p_i)^{2+k}\left(\sum_{l=1}^{i-1}p_l\right)^b \label{finform}
\end{align}

When $i = j = 1$, we simply have
\begin{align}
& \text{Pr}[Y_m = 1, Y_{m-1} = 1]\nonumber \\
&\,\,=\,\, \text{Pr}[X_1 \geq m+1] \nonumber\\
&\,\, = \sum_{k=0}^{K-1-m} \text{Pr}[X_1 = j+1+k] \nonumber \\
& \,\, = \sum_{k=0}^{K-1-m} \binom{K}{m+1+k} (p_1)^{m+1+k}(1-p_1)^{K-m-1-k} \label{anotherone4tr}
\end{align}

Finally, we compute $\text{Pr}[Y_m = i]$ as 
\begin{align}
&\text{Pr}[Y_m = i] \nonumber \\
&\,\, = \sum_{j=1}^{N} \text{Pr}[Y_m = i, Y_{m-1}=j] \nonumber \\
&\,\, =  \sum_{j = 1}^{i-1} \text{Pr}[Y-m = i, Y-{m-1} = j] \nonumber \\
 &\quad +\,\, \text{Pr}[Y_m = i, Y_{m-1} = i]\nonumber \\
&\,\,=\, \binom{K}{K-m}\left(  \left(\sum_{l=i}^{N}p_l \right)^{K-m} - \left(\sum_{l=i}^{N}p_l \right)^{K-m}  \right) \nonumber \\
 &\quad \left(\sum_{j=1}^{i-1}\left(\sum_{l=1}^{j}p_l\right)^m-\left(\sum_{l=1}^{j-1}p_l\right)^m\right) \nonumber \\
&\quad+ \,\, \text{Pr}[Y_m = i, Y_{m-1} = i] \nonumber \\
&\,\, =\, \binom{K}{K-m}\left(  \left(\sum_{l=i}^{N}p_l \right)^{K-m} - \left(\sum_{l=i+1}^{N}p_l \right)^{K-m}  \right) \nonumber \\
&\quad\left( \left(\sum_{l=1}^{i-1}p_l\right)^m\right) + \, \text{Pr}[Y_m = i, Y_{m-1} = i] \label{doneyet} 
\end{align}

Combing \eqref{doneyet} with \eqref{finform} and \eqref{doneyet} yields the desired result. This completes the proof.

\subsection{Proof of Proposition \ref{NUPobjfunc}} 
\label{C}
We wish to show that
\begin{multline}
 \mathbb{E}\left[\sum_{\mathcal{S}\in \mathcal{P}(\mathcal{U})\setminus \emptyset}\max_{k \in \mathcal{S}}\{|W_{S\setminus\{k\}}^{(d_k)}|\} \right] \\
 = \sum_{j=1}^{K-1} \sum_{i=0}^{K-1} \sum_{l=1}^{N} \binom{K-1-i}{j}\text{Pr}[Y_i=l]v_{l,j} \\
 +\sum_{i=0}^{K-1} \sum_{l=1}^{N} \text{Pr}[Y_{K-i-1} = l]v_{l,0} \label{restatement}
\end{multline}
if the memory inequality condition holds for the $v_{l,j}$ variables. We begin with an examination of the left hand side of the equation. Inside the expectation, we sum over all subsets $\mathcal{S}$ of the set of users $\mathcal{U}$. This can be rewritten as a double summation: in the inner summation, we sum over all subsets of size $j+1$, and in the outer summation, we sum over all $j$ from 0 to $K-1$, giving
 %\begin{align}
 %&\sum_{\mathcal{S}\in \mathcal{P}(\mathcal{U})\setminus \emptyset}\max_{k \in \mathcal{S}}\{|W_{S%\setminus\{k\}}^{(d_k)}|\}  \nonumber \\
 %=& \sum_{j=0}^{K-1} \sum_{\mathcal{S}\in \mathcal{P}(\mathcal{U})\setminus \emptyset : |\mathcal{S}| = j%+1}\max_{k \in \mathcal{S}}\{|W_{S\setminus\{k\}}^{(d_k)}|\}  \label{anotherstep}
 %\end{align}
  \begin{multline}
 \sum_{\mathcal{S}\in \mathcal{P}(\mathcal{U})\setminus \emptyset}\max_{k \in \mathcal{S}}\{|W_{S\setminus\{k\}}^{(d_k)}|\}  \\
 = \sum_{j=0}^{K-1} \sum_{\mathcal{S}\in \mathcal{P}(\mathcal{U})\setminus \emptyset : |\mathcal{S}| = j+1}\max_{k \in \mathcal{S}}\{|W_{S\setminus\{k\}}^{(d_k)}|\}  \label{anotherstep}
 \end{multline}
 
Replacing the $|W_\mathcal{S}|$ variables with the appropriate $v_{l,j}$ variables, \eqref{anotherstep} becomes
\begin{equation}
\sum_{j=0}^{K-1} \sum_{\mathcal{S}\in \mathcal{P}(\mathcal{U})\setminus \emptyset : |\mathcal{S}| = j+1}\max_{k \in \mathcal{S}}\{v_{d_k,j}\}. \label{furtherstill}
\end{equation}
For a fixed $j\geq 1$, we note that we send one transmission to each of the $\binom{K}{j+1}$ subsets of size $j+1$. For a fixed $\mathbf{d}$, let $k_i$ denote the user requesting the $i$-th most popular file, i.e. the file $i$-th smallest index. Then $k_1$ has requested the most popular file, and so by the memory inequality \eqref{3rdfifthconstraint}, $v_{d_{k_1},j}$ is the largest variable for any transmission to a subset of which $k_1$ is a member. Since $k_1$ is a member of $\binom{K-1}{j}$ subsets of size $j+1$ that contain $k_1$ as a member, the inner summation of \eqref{furtherstill} will have $\binom{K-1}{j}$ terms with the value $v_{d_{k_1},j}$. Similarly, user $k_2$ has requested the second most popular file, and so $v_{d_{k_2},j}$ will be the largest subfile for all subsets that contain $k_2$ but don't contain $k_1$. This constitutes $\binom{K-2}{j}$ subsets of size $j+1$.

This reasoning can be extended until all subsets are characterized in terms of their maximum $v_{l,j}$ variable. User $k_i$ requests the $i$-th most popular file, and so $v_{d_{k_i},j}$ will be the largest element sent in any subset containing $k_i$ but not containing $k_1, k_2, \dots, k_{i-1}$. Since there are $K-i$ users who are not users $k_1, \dots, k_i$, and user $k_i$ is already in the subset, there are $\binom{K-i}{j}$ subsets that contain $k_i$ but not $k_1, k_2, \dots, k_{i-1}$. We can therefore eliminate the $\max\{\}$ term from the inner sum of \eqref{furtherstill} to obtain, for $j = 1, \dots, K-1$, 
\begin{equation}
\sum_{i = 1}^{K} \binom{K-i}{j} v_{d_{k_i},j}.   \label{almost there}
\end{equation}
As noted earlier, the memory inequality reverses for $j=0$, so the least popular files take up the most memory in that case; the reasoning is the same as in the above, but we instead obtain
\begin{equation}
\sum_{i = 1}^K \binom{K-i}{0} v_{d_{k_{K+1-i}},0} = \sum_{i = 1}^K v_{d_{k_{K+1-i}},0}, \label{almooost}
\end{equation}

All that remains is to compute the expectation of these terms with respect to the demand vectors. Using the linearity of expectation and the results of \eqref{anotherstep}-\eqref{almooost}, the left hand side of \eqref{restatement} reduces to
\begin{eqnarray}
\sum_{j=1}^{K-1} \sum_{i = 1}^{K} \binom{K-i}{j} \mathbb{E}[v_{d_{k_i},j}] + \sum_{i = 1}^{K} \mathbb{E}[v_{d_{k_{K+1-i}},0}] \label{neatobureato}
\end{eqnarray}

To compute the expected value of the $v_{d_{k_i},j}$ variables ($j = 1, \dots, K-1$), we note that it has $N$ possible values, $v_{1,j}, v_{2,j}, \dots, v_{N,j}$, and the probability of each outcome can be obtained from Lemma \ref{problemma} in the following way. We have $v_{d_{k_i},j} = v_{l,j}$ if $l$ is the $i$-th most popular file in the request vector $\mathbf{d}$; since the files are labelled in terms of decreasing order of popularity, the $i$-th most popular file requested is represented by the $i$-th smallest index in $\mathbf{d}$. Thus the probability that $d_{k_i} = l$ is equivalent to the probability that $l$ is the $i$-th smallest index in $\mathbf{d}$, and so by Lemma \ref{problemma}, we have
\begin{eqnarray}
\mathbb{E}[v_{d_{k_i},j}] & = & \sum_{l=1}^{N} \text{Pr}[d_{k_i} = l]v_{l,j} \nonumber \\
				       & = & \sum_{l=1}^{N} \text{Pr}[Y_{i-1} = l]v_{l,j}. \label{expeccomp} 
\end{eqnarray}
For $j = 0$, the size ordering is reversed, so we are concerned with the \emph{largest} indices of $\mathbf{d}$. However, as has been noted already, the $i$-th largest index of $\mathbf{d}$ must necessarily be the $K+1-i$-th smallest index of $\mathbf{d}$, which gives
%Lemma \ref{problemma} can also be used to compute this distribution: to compute the probability distribution of the $i$-th largest element of $\mathbf{d}$, form an auxiliary variable $\tilde{\mathbf{d}}$ identical to $\mathbf{d}$ but with the labels in reverse order. Then the probability distribution of the $i$-th smallest element of $\tilde{\mathbf{d}}$, which can be computed with Lemma \ref{problemma}, is equivalent to the probability distribution of the $i$-th largest element of $\mathbf{d}$, which we denote by $\text{Pr}[\tilde{Y}_{i+1} = l]$. We can then write
\begin{eqnarray}
\mathbb{E}[v_{d_{k_{K+1-i}},0}] & = & \sum_{l=1}^{N} \text{Pr}[d_{k_{K+1-i}} = l]v_{l,0} \nonumber \\
				       & = & \sum_{l=1}^{N} \text{Pr}[Y_{K-i} = l]v_{l,0}. \label{expeccomp2} 
\end{eqnarray}
Combing \eqref{expeccomp}-\eqref{expeccomp2}, we see that \eqref{neatobureato} is equal to
\begin{multline}
\sum_{j=1}^{K-1} \sum_{i = 1}^{K} \binom{K-i}{j}  \sum_{l=1}^{N} \text{Pr}[Y_{i-1} = l]v_{l,j} \\
+ \sum_{i = 1}^{K} \sum_{l=1}^{N} \text{Pr}[Y_{K-i} = l]v_{l,0}. \label{finalexpr4this}
\end{multline}
We complete the proof through a cosmetic change of variables $i' = i - 1$ to obtain the desired expression on the right-hand side of \eqref{restatement}.

\subsection{Proof of Proposition \ref{NUMprop}} 
\label{D}
We follow reasoning similar to what we have already seen in 
%Section \ref{NUMPF}
the previous proof, where users are divided into subsets that require the same amount of data to be sent to them, and then count how many such subsets there are. For this proof, however, we instead divide the various subsets into subsets containing only small-cache users, subsets containing only large-cache users, and subsets containing both large- and small-cache users. 

But first, we note that since the files are all the same size and length, the transmission length will be independent of the request vector $\mathbf{d}$, and so we have 
\begin{multline}
\mathbb{E}\left[\sum_{\mathcal{S}\in \mathcal{P}(\mathcal{U})\setminus \emptyset}\max_{k \in \mathcal{S}}\{|W_{S\setminus\{k\}}^{(d_k)}|\} \right]  \\
=  \sum_{\mathcal{S}\in \mathcal{P}(\mathcal{U})\setminus \emptyset}\max_{k \in \mathcal{S}}\{|W_{S\setminus\{k\}}^{(d_k)}|\}. \label{onemoreeq}
\end{multline}

As discussed above, the sum in the above expression is over all $\mathcal{S}\in \mathcal{P}(\mathcal{U})\setminus \emptyset$, which we can separate into small, large, and mixed sets. For a fixed subset size of $j+1$, there are $\binom{K_S}{j+1}$ sets of small users, $\binom{K_L}{j+1}$ sets of large users, and 
\begin{equation}
\sum_{i = 1}^j \binom{K_S}{i}\binom{K_L}{j+1-i} 
\end{equation}
groups of at least one small user and at least one large user. For a set of $j+1$ small users, every subfile in a single coded transmission is cached by $j$ small users, and so has the size $v_{j,S}$. Similarly, for any set of $j+1$ large users, the transmission has the size $v_{j,L}$.

For the mixed subset case, we must consider three cases. First, when there are at least 2 small users and 2 large users in the subset of $j+1$ users, then since every subfile sent is cached on $j$ of the $j+1$ users, there must be at least 1 small user and 1 large users among those $j$ users,  and so every subfile must be of size $v_{j,M}$. However, if there is only one small user in the subset of $j+1$ users, then the subfile requested by the small user will have been stored on the caches of $j$ large users, and so will have size $v_{j,L}$. The length of the entire transmission will therefore also be of size $v_{j,L}$. The third case occurs when there is only one large users in the subset of $j+1$ users. Then the subfile requested by the large user will be store on the caches of $j$ small user and so will be of size $v_{j,S}$, while every other subfile is cached on a mixed set of $j$ users and so will be of size $v_{j,M}$; the entire transmission will therefore be of length $v_{j,M}$. \footnote{In the case where a subset of size 2 contains one large user and one small user, obviously the entire transmission is of length $v_{1,L}$.}

So, in addition to the $\binom{K_L}{j+1}$ transmissions of size $v_{j,L}$ sent for groups entirely consisting of entirely large users, there are $\binom{K_S}{1}\binom{K_L}{j}$ transmissions of the same size for those mixed subsets with only one small user. The total number of transmissions of size $v_{j,M}$ can then be simplified using Lemma \ref{CVC} as
\begin{align}
&\sum_{i=2}^j \binom{K_S}{i}\binom{K_L}{j+1-i} \nonumber \\
 &\,\, = \sum_{i=0}^{j+1} \binom{K_S}{i}\binom{K_L}{j+1-i} - \binom{K_S}{j+1} \nonumber \\
 &\quad - \binom{K_S}{1}\binom{K_L}{j} - \binom{K_L}{j+1} \nonumber \\
 &\,\, = \binom{K}{j+1} - \binom{K_S}{j+1} \nonumber \\
 &\quad - \binom{K_S}{1}\binom{K_L}{j} - \binom{K_L}{j+1}
\end{align}

Altogether, \eqref{onemoreeq} reduces to
\begin{multline}
 \sum_{j=0}^{K-1} \binom{K_s}{j+1}(v_{j,S}-v_{j,M}) + \binom{K}{j+1}v_{j,M} \\
 + \left(\binom{K_L}{j+1} +\binom{K_S}{1}\binom{K_L}{j}\right)(v_{j,L}-v_{j,M}), \nonumber
 \end{multline}
which is what we aimed to show. We make a special note that the formula is indeed sensible for $j = 0$: the $j = 0$ term reduces to $\binom{K}{1}v_{0,M} = Kv_{0}$, as needed for the individual transmissions to the K users.

\subsection{Proof of Proposition \ref{finprop}}
\label{E}
We derive the terms of \eqref{finobjfunc} in the order that they appear. In general, we do this using the following steps. First, we identify a certain group of subsets that have similar user composition; then for that group, we determine the number of transmissions that the largest subfile will be in, the number of transmissions that the second largest subfile will be in, and so on. Finally, we compute the expected size of the maximum subfile, the second largest subfile, and so on. This approach will be familiar from previous proofs, but we nevertheless repeat it here due to the complexity of \eqref{finobjfunc}.

The groups of subsets that the seven terms of \eqref{finobjfunc} correspond to are, in order: subsets of size one, subsets of size greater than one containing only large users, subsets of size greater than one containing only small users, mixed subsets containing more than one user but only one small user, mixed subsets containing only one large user but more than one small user, subsets containing greater than or equal to $\max\{K_S,K_L\}+2$ users, and subsets containing at least two small and two large users that are less than $\max\{K_S,K_L\}+2$ users. We label these sets of subsets $\mathcal{S}_1, \dots, \mathcal{S}_7$ respectively. The following lemma shows these sets form a partition (in the loose sense of the word discussed earlier) of $\mathcal{P}(\mathcal{U})\setminus \emptyset$, and so the sum over all $\mathcal{S} \in \mathcal{P}(\mathcal{U})\setminus \emptyset$ at the beginning of \eqref{finobjfunc} can equivalently be done over all subsets in $\mathcal{S}_1$, then all subsets in $\mathcal{S}_2$, and so on, so that all subsets of users in $\mathcal{P}(\mathcal{U})\setminus \emptyset$ will have been accounted for precisely once.
\begin{lemma}
\label{Elemma1}
For the sets $\mathcal{S}_1, \dots, \mathcal{S}_7$ described above,
\begin{equation}
\mathcal{P}(\mathcal{U})\setminus \emptyset = \bigcup_{i=1}^{7} \mathcal{S}_i,
\end{equation}
and the $\mathcal{S}_i$ are mutually disjoint. 
\end{lemma}
\emph{Proof:} That $\bigcup_{i=1}^{7} \mathcal{S}_i \subseteq \mathcal{P}(\mathcal{U})\setminus \emptyset$ is trivial: for any i, any set in $\mathcal{S}_i$ is a non-empty subset of users, and so must be contained in $\mathcal{P}(\mathcal{U})\setminus \emptyset$. To show that $\mathcal{P}(\mathcal{U})\setminus \emptyset \subseteq \bigcup_{i=1}^{7} \mathcal{S}_i$, consider the number of users in an arbitrary subset of users $\mathcal{S} \in \mathcal{P}(\mathcal{U})\setminus \emptyset$: if it is one, then $\mathcal{S} \subseteq \mathcal{S}_1 $ and if it is greater than or equal to $\max\{K_S,K_L\}+2$, then it must be mixed because there are not enough of any one type of user to comprise the entire group, and so is must be that $\mathcal{S} \subseteq \mathcal{S}_6$. Otherwise, suppose $1 < |\mathcal{S}| < \max\{K_S,K_L\}+1$, consider the number of small users, $k_S$, in $\mathcal{S}$. If $k_S$ = 0, then $\mathcal{S}$ is contains only large users, and so $\mathcal{S} \subseteq \mathcal{S}_2$. If $k_S = 1 $, then we have $\mathcal{S} \subseteq \mathcal{S}_4$. If $1<k_S < |\mathcal{S}|$, then either the number of large users is either one, or more than one; if it is one, then $\mathcal{S} \subseteq \mathcal{S}_5$, while if it is more than one, then $\mathcal{S} \subseteq \mathcal{S}_7$. Finally, if $k_S = |\mathcal{S}|$, there are only small users, and so $\mathcal{S} \subseteq \mathcal{S}_3$, proving that indeed $\mathcal{P}(\mathcal{U})\setminus \emptyset \subseteq \bigcup_{i=1}^{7} \mathcal{S}_i$. The mutual disjointedness is obvious once it is noted that a subset containing only one large/small user or no large/small users cannot exceed a size of $\max\{K_S,K_L\}$ or $\max\{K_S,K_L\}+1$ respectively. Each $\mathcal{S} \subseteq \mathcal{P}(\mathcal{U})\setminus \emptyset $ thus falls into one and only one set $\mathcal{S}_i$, proving the lemma.

We remark before continuing that, given the specific values of $K_S, K_L$, some of the above subsets may be empty. As per the notation adopted in this paper, a sum over an empty set is identically zero, and so this will not affect our subsequent calculations. In terms of the expressions below, this will correspond to binomial coefficients $\binom{n}{k}$ with $n < 0$ or $k > n$, both of which, by our notation, gives $\binom{n}{k} = 0$.

So per the above discussion, we can change the summation over all subsets of $\mathcal{P}(\mathcal{U})\setminus \emptyset $ into seven summations over one of the $\mathcal{S}_i$ each:
\begin{multline}
\mathbb{E}\left[\sum_{\mathcal{S}\in \mathcal{P}(\mathcal{U})\setminus \emptyset}\max_{k \in \mathcal{S}}\{|W_{\mathcal{S}\setminus\{k\}}^{(d_k)}|\} \right]  \\
 =\sum_{i=1}^{7}\sum_{\mathcal{S} \in \mathcal{S}_i} \mathbb{E}\left[ \max_{k \in \mathcal{S}}\{ |W^{(d_k)} _{\mathcal{S}\setminus\{k\}}| \} \right] \label{SumS}
\end{multline} 
This allows us to analyze each subset of subsets separately.

We begin with the analysis of $\mathcal{S}_1$, i.e. to broadcasts of individual users. Since each transmission is to only one person, we get
\begin{align}
\sum_{\mathcal{S} \in \mathcal{S}_1} \mathbb{E}\left[ \max_{k \in \mathcal{S}}\{ |W^{(d_k)} _{\emptyset}| \} \right] = \sum_{k = 1}^{K}\mathbb{E}\left[ |W^{(d_k)}_{\emptyset}| \right] 
 = \sum_{k = 1}^{K}\mathbb{E}\left[ v_{d_k,0} \right] \nonumber 
\end{align}
The above sum is over all users from $k = 1$ to $k= K$, i.e. in lexicographic order. But we can instead sum over all users by adding the user requesting the largest subfile, then the user requesting the second largest subfile, and so on. Using the index $i$ to indicate the user requesting the $(i+1)$-th largest subfile, we can write the expectation $\mathbb{E}[v_{d_k,0}]$ in terms of the random variable $Y_{i}$ as defined in Lemma \ref{problemma} to obtain
\begin{align}
 \sum_{k = 1}^{K}\mathbb{E}\left[ v_{d_k,0} \right] = \sum_{i = 0}^{K-1} \mathbb{E}\left[ v_{f_d(i+1),0} \right] 
= \sum_{i = 0}^{K-1} \sum_{l = 1}^N  \text{Pr}[Y_i = l]v_{l,0}, \label{S1}
\end{align} 
which is the first term of \eqref{finobjfunc}, with $v_{l,0} = v_{l,0}^M$ as per constraint \eqref{zeroeq}. Here, $f_d(i)$ denotes the index of the $i$-th largest file in the request vector $\mathbf{d}$ (recall that $f(i)$ was used earlier to denote the $i$-th largest file in the set of all files).

We next consider $\mathcal{S}_2$, the set of user subsets with more than one user containing only large-cache users. There are $\binom{K_L}{j}$ user subsets of size $j+1$ in $\mathcal{S}_2$, for $j$ values ranging from 1 to $K_L-1$; we cannot have a subset of only large users that contains more members than there are large users. Since there are only large users in these subsets, the subfiles sent will stored on $j$ large users caches, and so only subfiles of size $v_{j,l}^L$ are sent. As we saw in earlier proofs, the largest subfile requested (i.e. corresponding to the file with the smallest index), will be sent to $\binom{K_L-1}{j}$ subsets, the second largest subfile is the largest subfile for $\binom{K_L-2}{j}$ subsets, and in general, the $i$-th largest subfile sent will be sent in $\binom{K_L-i}{j}$ subsets, giving
\begin{align}
&\sum_{\mathcal{S} \in \mathcal{S}_2} \mathbb{E}\left[ \max_{k \in \mathcal{S}}\{ |W^{(d_k)} _{\mathcal{S}\setminus\{k\}}| \} \right] \nonumber \\
&\,\,= \sum_{j = 1}^{K_L-1} \sum_{i=1}^{K_L} \binom{K_L-i}{j} \mathbb{E}\left[ v_{f^L_d(i),j}^L \right] \nonumber \\
&\,\, = \sum_{j = 1}^{K_L-1} \sum_{i=1}^{K_L} \binom{K_L-i}{j} \sum_{l=1}^N \text{Pr}[Y^L_{i-1} = l]v_{l,j}^L \nonumber \\
&\,\, = \sum_{j = 1}^{K_L-1} \sum_{i=0}^{K_L-1} \sum_{l=1}^N \binom{K_L-i+1}{j} \text{Pr}[Y^L_{i} = l]v_{l,j}^L, \label{S2}
\end{align} 
where the last line is obtained by rearranging the terms and using a minor change of variable for the index of summation $i$. This is the second term of \eqref{finobjfunc}. Here we use $f^L_{d}(i)$ to refer to the $i$-th largest file requested within the set of large users, and by $\text{Pr}[Y^L_i = l]$, we mean the probability that file $l$ is the $i+1$th largest file requested within the set of large users. We can compute $\text{Pr}[Y^L_i = l]$ using Lemma \ref{problemma} with $N$ files (outcomes) and $K_L$ users (trials). The change from the second to third lines above then follows immediately from the definition of expectation (see Appendix \ref{C}). 

Using identical reasoning for $\mathcal{S}_3$, the set of user subsets of size greater than 1 with only small users, we can obtain
\begin{align}
&\sum_{\mathcal{S} \in \mathcal{S}_3} \mathbb{E}\left[ \max_{k \in \mathcal{S}}\{ |W^{(d_k)} _{\mathcal{S}\setminus\{k\}}| \} \right] \nonumber \\
&\,\,= \sum_{j=1}^{K_S-1} \sum_{i = 0}^{K_S-1}\sum_{l=1}^N \binom{K_S-i+1}{j}\text{Pr}[Y_i^S = l]v_{l,j}^S \label{S3}
\end{align}
which is the third term of \eqref{finobjfunc}. Here, $\text{Pr}[Y^S_i = l]$ is the probability that file $l$ is the $i+1$-th largest file requested among all small users. This can also be computed using Lemma \ref{problemma}, but with $N$ outcomes and $K_S$ trials.

Next, we consider $\mathcal{S}_4$, the set of mixed subsets containing more than one user but only one small user. We saw in the non-uniform cache memory case that the coded transmissions to these kinds of groups will consist almost entirely of subfiles whose size is described by mixed variables $v_{l,j}^M$, because the subfiles are stored on a mixed subset of users' caches, save for one subfile whose size is described by a large variable $v_{l,j}^L$, because that subfile is stored only on large user caches. Due the memory inequality constraints \eqref{7thfinconstraintp1}-\eqref{7thfinconstraintp3} that prioritize cache size over file size, the one large variable (corresponding to the file requested by the one small user) will necessarily be the maximum value. 

The size of the transmissions sent to these kinds of subsets will therefore depend on what files are requested by the small-cache users. Each small cache user will be in $\binom{K_L}{j}$ many of these subsets for a subset size of $j+1$, where $j$ takes values from 1 to $K_L$; if $j$ was any larger, there would have to be more than one small user.  We therefore have
\begin{align}
\sum_{\mathcal{S} \in \mathcal{S}_4} \mathbb{E}\left[ \max_{k \in \mathcal{S}}\{ |W^{(d_k)} _{\mathcal{S}\setminus\{k\}}| \} \right]  =  \sum_{j = 1}^{K_L} \sum_{i=1}^{K_S} \binom{K_L}{j} \mathbb{E}\left[ v_{f^S_d(i),j}^L \right] \label{interstep}
\end{align} 
We use $f^S_d(i)$ to denote the index of the $i$-th largest file requested by a small-user. Consequently, the second sum in \eqref{interstep} is over all small cache users, in decreasing order of the file size they requested. This allows us to compute the expectation in \eqref{interstep} using the $Y_i^S$ variables in the following way:
\begin{align}
&\sum_{\mathcal{S} \in \mathcal{S}_4} \mathbb{E}\left[ \max_{k \in \mathcal{S}}\{ |W^{(d_k)} _{\mathcal{S}\setminus\{k\}}| \} \right] \nonumber \\
&\,\,=  \sum_{j = 1}^{K_L} \sum_{i=1}^{K_S} \binom{K_L}{j} \sum_{l=1}^N \text{Pr}[Y^S_{i-1} = l]v_{l,j}^L \nonumber \\
&\,\,=  \sum_{j = 1}^{K_L} \sum_{i=0}^{K_S-1}\sum_{l=1}^N \binom{K_L}{j}\text{Pr}[Y^S_{i} = l]v_{l,j}^L. \label{S4}
\end{align}
The final step is once again attained with a rearranging of terms and a change of variable for the $i$ index of summation. This gives the fourth term in \eqref{finobjfunc}.

The fifth term is obtained using similar reasoning. This term corresponds to $\mathcal{S}_5$, the set of mixed user subsets containing exactly one large user and more than one small user. Here, the transmitted subfiles will all be stored on the caches of a mixed subset of users, except for the subfile requested by the large user, which will be stored on the caches of every other user in the subset, i.e. all small users. The large user's requested subfile will have a size described by a small variable $v_{l,j}^S$, and so due to the memory inequality constraints \eqref{7thfinconstraintp1}-\eqref{7thfinconstraintp3}, will never be the largest subfile transmitted; once again, it is the small-cache user requests that determine the largest subfile. The largest subfile requested by a small user will be transmitted to $\binom{K_S-1}{j-1}\binom{K_L}{1}$ subsets of size $j+1$; the second largest subfile requested among small users will be the largest subfile transmitted when the largest subfile requested is not also being transmitted to that subset, and so will be transmitted $\binom{K_S-2}{j-1}\binom{K_L}{1}$ times. In general, the $i$-th largest subfile requested among small users will be transmitted only when the previous $i-1$ largest subfiles are not also being transmitted, and so will be sent $\binom{K_S-i}{j-1}\binom{K_L}{1}$ times. 

There are subsets of size 3 through $K_S+1$ in $\mathcal{S}_5$, so indexing subset size with $j+1$ yields 
\begin{align}
&\sum_{\mathcal{S} \in \mathcal{S}_5} \mathbb{E}\left[ \max_{k \in \mathcal{S}}\{ |W^{(d_k)} _{\mathcal{S}\setminus\{k\}}| \} \right] \nonumber \\
&\,\,=  \sum_{j = 2}^{K_S} \sum_{i=1}^{K_S-1} \binom{K_S-i}{j-1}\binom{K_L}{1} \mathbb{E}[v^M_{f^S_d(i),j}] \nonumber \\
&\,\,=  \sum_{j = 2}^{K_S} \sum_{i=1}^{K_S-1}\binom{K_S-i}{j-1}\binom{K_L}{1}\left(\sum_{l=1}^N \text{Pr}[Y^S_{i-1} = l]v_{l,j}^M\right) \nonumber \\
&\,\,=  \sum_{j = 2}^{K_S} \sum_{i=0}^{K_S-2}\sum_{l=1}^N \binom{K_S-1-i}{j-1}\binom{K_L}{1}\text{Pr}[Y^S_{i} = l]v_{l,j}^M. \label{S5}
\end{align}
The last line is once again obtained through rearranging terms and doing a change of variables for the index of summation $i$. This is the fifth term of \eqref{finobjfunc}.

The sixth term of \eqref{finobjfunc} contains the terms for $S_6$, the set of user subsets with more than $\max\{K_S,K_L\}+1$ users. These subsets are precisely large enough that they are all mixed and have at least two of each user type in them. Thus only subfiles stored on the caches of mixed subsets of users will be sent, and so they will have a size given by a mixed variable $v_{l,j}^M$. The only factor that determines the largest subfile for a given subset will therefore be the file index. There are no restrictions on subset composition, and so we find ourselves in a familiar situation: the largest subfile requested will be the largest file sent for $\binom{K-1}{j}$ subsets, the second largest subfile requested will be the largest subfile sent for $\binom{K-2}{j}$ subsets, and so on, such that the $i$-th largest subfile is the largest subfile sent for $\binom{K-i}{j}$ subsets. This gives
\begin{align}
&\sum_{\mathcal{S} \in \mathcal{S}_6} \mathbb{E}\left[ \max_{k \in \mathcal{S}}\{ |W^{(d_k)} _{\mathcal{S}\setminus\{k\}}| \} \right] \nonumber \\
&\,\,=  \sum_{j = \max\{K_S,K_L\}+1}^{K-1} \sum_{i=1}^{K} \binom{K-1-i}{j}\mathbb{E}\left[v^M_{f_d(i),j}\right] \nonumber \\
&\,\, =  \sum_{j = \max\{K_S,K_L\}+1}^{K-1} \sum_{i=0}^{K-1} \binom{K-1-i}{j}\text{Pr}[Y_i = l]v^M_{l,j}, 
\label{S6}
\end{align}
the sixth term of \eqref{finobjfunc}.

The seventh and final term of \eqref{finobjfunc} is by far the most complicated term. It corresponds to $\mathcal{S}_7$, the set of subsets with at least two large-cache and two small-cache users, but less than $\max\{K_S,K_L\}+2$ users. Here, every subfile sent will be stored on the caches of a mixed subset of users and so will have a size given by a mixed variable $v_{l,j}^M$. The difficulty arises when we try to characterize the number of transmissions for each of the largest subfile requested, second largest subfile requested, and so on - these numbers are dependent on whether the file was requested by a large user or by a small user. An example will illustrate this fact: consider the third largest file requested among all users and transmissions to subsets of 5 users. If a large-cache user has requested the largest subfile, and another large-cache user requests the second largest subfile, then the number of transmissions where the third largest subfile requested is the largest subfile transmitted is $\sum_{n=1}^{2} \binom{K_L-3}{n}\binom{K_S}{4-n}$ if the subfile is requested by a large-cache user, and $\sum_{n=2}^{3}\binom{K_L-2}{n}\binom{K_S-1}{4-n}$ if it is requested by a small-cache user. These two number are clearly not equal in general, and so the cache size of the user making the request matters.

%Nevertheless, it is possible to compute the expected rate with a number of terms that scales as a polynomial function of $N$ and $K$. To this end, let $R_{\mathbf{d}}^{\mathcal{S}_i}$ denote the number of bits sent to subsets in $\mathcal{S}_i$ as part of the transmissions required to satisfy the request $\mathbf{d}$, i.e. such that $R_{\mathbf{d}}=\sum_{i=1}^7 R_{\mathbf{d}}^{\mathcal{S}_i}$. Let $R_{\mathbf{d},j}^{\mathcal{S}_i}$ be number of bits of $R_{\mathbf{d}}^{\mathcal{S}_i}$ sent to subsets of size $j+1$ such that $R_{\mathbf{d}}^{\mathcal{S}_i} = \sum_{j=0}^{K-1} R_{\mathbf{d},j}^{\mathcal{S}_i} $; some of the summands here may be zero depending on which $\mathcal{S}_i$ is chosen. Finally, let $R_{\mathbf{d},j}^{\mathcal{S}_i}(n)$ be the number of bits in the previous that are sent when the $n$-th largest file requested is the largest file transmitted in the subset, i.e. such that $R_{\mathbf{d},j}^{\mathcal{S}_i} = \sum_{n=1}^K R_{\mathbf{d},j}^{\mathcal{S}_i}(n)$. These values have been implicitly computed for $\mathcal{S}_1$ through $\mathcal{S}_6$ earlier in this appendix; we only introduce this opaque notation now because the complexity of the accounting done for $\mathcal{S}_7$ demands it. 

Nevertheless, it is possible to compute the expected rate with a number of terms that scales as a polynomial function of $N$ and $K$. To this end, let $R_{\mathbf{d},j}^{\mathcal{S}_i}(n)$ denote the number of bits sent to subsets in $\mathcal{S}_i$ of size $j+1$ as part of the transmissions required to satisfy the request $\mathbf{d}$, when the $n$-th largest file in $\mathbf{d}$ is the largest subfile transmitted in that subset. By this definition, we have $R_{\mathbf{d}}=\sum_{i=1}^7 \sum_{j=0}^{K-1} \sum_{n=1}^K R_{\mathbf{d},j}^{\mathcal{S}_i}(n)$. These values have been implicitly computed for $\mathcal{S}_1$ through $\mathcal{S}_6$ earlier in this appendix; we only introduce this opaque notation now because the complexity of the accounting done for $\mathcal{S}_7$ demands it.

The $R_{\mathbf{d},j}^{\mathcal{S}_7}(n)$ quantity can be further decomposed: the number of bits sent in this case is equal to the product of the number of transmission sent in this case, denoted by $T(n)$, and the number of bits transmitted per transmission, which is given buy the appropriate $v_{l,j}^M$ variable. We can then compute the $\mathcal{S}_7$ term of \eqref{finobjfunc} using conditional expectation in the following way:
\begin{align}
\sum_{\mathcal{S} \in \mathcal{S}_7} \mathbb{E}\left[ \max_{k \in \mathcal{S}}\{ |W^{(d_k)} _{\mathcal{S}\setminus\{k\}}| \} \right] =&\,\,  \mathbb{E}\left[ \sum_{j=3}^{\max\{K_S,K_L\}}\sum_{n=1}^K R_{\mathbf{d},j}^{\mathcal{S}_7}(n) \right] \nonumber \\
=& \sum_{j=3}^{\max\{K_S,K_L\}}\sum_{n=1}^K \mathbb{E}\left[R_{\mathbf{d},j}^{\mathcal{S}_7}(n) \right]  \label{prevugh}
\end{align}
For a fixed $j$ value, we compute the expectation conditional on the fact that the $n$-th largest file is file $l$, i.e. $Y_{n-1} = l$:
\begin{align}
&\sum_{n=1}^K \mathbb{E}\left[R_{\mathbf{d},j}^{\mathcal{S}_7}(n) \right] \nonumber \\
&\,\,= \sum_{n=1}^K\sum_{l=1}^N \mathbb{E}\left[R_{\mathbf{d},j}^{\mathcal{S}_7}(n) | Y_{n-1}=l \right]\text{Pr}[Y_{n-1}=l] \nonumber \\
&\,\,= \sum_{n=1}^K\sum_{l=1}^N \mathbb{E}\left[v_{f_{d}(n),j}^MT(n) | Y_{n-1}=l \right]\text{Pr}[Y_{n-1}=l] \nonumber \\
&\,\, = \sum_{n=1}^K\sum_{l=1}^N v_{l,j}^M\mathbb{E}\left[T(n) | Y_{n-1}=l \right]\text{Pr}[Y_{n-1}=l]. \label{ugh}
\end{align}
The last line \eqref{ugh} is obtained because, given that $Y_{n-1}=l$, it follows immediately that $f_d(n) = l$, and so we have $v_{f_{d}(n),j}^M = v_{l,j}^M$, which is no longer a random quantity. 

Comparing \eqref{ugh} to the form of \eqref{finobjfunc} in the statement of the proposition, we see that all that remains is to show that $\mathbb{E}\left[T(n) | Y_{n-1}=l \right] = \nu(n,j)$. 
First, we note that $\mathbb{E}\left[T(n) | Y_{n-1}=l \right] = \mathbb{E}\left[T(n) \right]$, since the number of transmissions in which the $n$-th largest subfile requested is the largest subfile sent to the subset depends only on the index $n$ but not the identity of the n-th largest subfile. Next, letting $S(n)$ denote the number of small users in the set of users who requested the $n-1$ largest files we further decompose the expectation using conditional expectation:
\begin{align}
\mathbb{E}\left[  T(n) \right] 
= \sum_{m= 0 }^{n-1}\mathbb{E}\left[  T(n) | S(n) = m\right]\text{Pr}[S(n) = m] \label{ugh2}. 
\end{align}
 And further, if $D_S(n) = 1 $ represents the event that a small user requested the $n$-th largest file and $D_S(n) = 0$ representing the event that a large user did it, we have
 \begin{align}
 \mathbb{E}\left[  T(n) \right] 
&= \sum_{m= 0 }^{n-1}\mathbb{E}\left[  T(n) | S(n) = m\right]\text{Pr}[S(n) = m]  \nonumber \\
&= \sum_{m= 0 }^{n-1}\sum_{r = 0}^{1}\mathbb{E}\left[  T(n) | S(n) = m,D_S(n) = r\right]\nonumber\\
&\quad \text{Pr}[D_S(n) = r|S(n) = m]\text{Pr}[S(n) = m] \label{ugh3}
 \end{align}
Now with \eqref{ugh}-\eqref{ugh3}, we have finally expressed the original expectation in \eqref{prevugh} in terms of quantities that can be computed directly.
  
We begin with $\text{Pr}[S(n) = m]$, the probability that there are $m$ small-cache users in the set of users who have the $n-1$ largest files among all files requested. Since all users have the same preferences, these probabilities are simply determined by the relative numbers of large and small users. Indeed $S(n)$ has a hypergeometric distribution: the probability that $m$ of the $n-1$ largest files requested are requested by small users (and thus $n-1-m$ of these files are requested by large users) is given by
\begin{equation}
\text{Pr}[S(n) = m] = \frac{\binom{K_S}{m}\binom{K_L}{n-1-m}}{\binom{K}{n-1}} \label{hyperg}.
\end{equation}
Next we consider $\text{Pr}[D_S(n) = r|S(n)=m]$, which is obtained with similar reasoning. Once again, since the large and small users have the same preferences, only their relative numbers will determine the probabilities. Since $n-1$ users have already been accounted for, there are $K-(n-1)$ users left to choose from, and if $m$ of them are small users, there are $K_S-m$ small users left and $K_L - (n-1-m)$ large users left. This gives
\begin{equation}
\text{Pr}[D_S(n) = 1|S(n)=m] = \frac{K_S-m}{K-n+1} \label{p2}
\end{equation}  
and
\begin{equation}
\text{Pr}[D_S(n) = 0|S(n)=m] = \frac{K_L-n+1+m}{K-n+1} \label{p3}
\end{equation}

Finally, we compute $\mathbb{E}\left[  T(n) | S(n) = m,D_S(n) = r\right]$; the number of transmissions $T(n)$ is deterministic given the values of $S(n)$ and $D_S(n)$, so no probabilities will be involved in the calculation.  First, if $r = 1$, i.e. a small user has the $n$-th largest file request. In this case, the corresponding subfile is the largest subfile transmitted for any transmission to a subset with at least one other small user and two large users, but not the users responsible for the $n-1$ larger requested files. This number is obtained as
\begin{multline}
 \mathbb{E}\left[  T(n) | S(n) = m,D_S(n) = 1\right]  \\
= \sum_{i = 1}^{j-2} \binom{K_S-m-1}{i}\binom{K_L-n+1+m}{j-i} \label{p4}
\end{multline}
for a subset size of $j+1$. The equivalent number for $r = 0$, i.e. a large user has the $n$-th largest file request, is
\begin{multline}
 \mathbb{E}\left[  T(n) | S(n) = m,D_S(n) = 0\right] \\
= \sum_{i = 2}^{j-1} \binom{K_S-m}{i}\binom{K_L-n+m}{j-i} \label{p5}.
\end{multline}

Substituting the expressions in \eqref{hyperg}-\eqref{p5} into the appropriate places in \eqref{ugh} - \eqref{ugh3} yields the desired term, i.e. the seventh and final term of \eqref{finobjfunc}, which concludes the proof.
%Bibliography

\bibliographystyle{IEEEtran}
\bibliography{IEEEabrv,FLCIOBib}

\end{document}